\documentclass[a4paper]{article}

\pdfoutput=1

\usepackage[margin=1in]{geometry}
\usepackage[utf8]{inputenc}
\usepackage{float}
\usepackage{latexsym, amsmath, amssymb, mathrsfs, graphicx}
\usepackage{empheq}
\usepackage{hyperref}
\usepackage{cite}
\usepackage{authblk}
\usepackage{tikz}
\usetikzlibrary{snakes}
\usetikzlibrary{decorations.pathreplacing}
\usetikzlibrary{plotmarks}
\newcommand {\be}{\begin{equation}}
\newcommand {\ee} {\end{equation}}
\newcommand {\bea}{\begin{eqnarray}}
\newcommand {\eea} {\end{eqnarray}}

\begin{document}

\title{Quasilocal charges and progress towards the complete GGE for field theories with non-diagonal scattering}
\author{Eric Vernier\footnote{evernier@sissa.it}\,  and Axel Cort\'{e}s Cubero\footnote{acortes@sissa.it}}
\affil{SISSA and INFN, via Bonomea 265, 34136 Trieste, Italy. }
\date{}							

\maketitle

\begin{abstract}
It has recently been shown that some integrable spin chains possess a set of quasilocal conserved charges, with the classic example being the spin-$\frac{1}{2}$ XXZ Heisenberg chain. These charges have been proven to be essential for properly describing stationary states after a quantum quench, and must be included in the generalized Gibbs ensemble (GGE). We find that similar charges are also necessary for the GGE description of integrable quantum field theories with non-diagonal scattering. A stationary state in a non-diagonal scattering theory is completely specified by fixing the mode-occupation density distributions of physical particles, as well auxiliary particles which carry no energy or momentum. We show that the set of conserved charges with integer Lorentz spin, related to the integrability of the model, are unable to fix the distributions of these auxiliary particles, since these charges can only fix kinematical properties of physical particles.  The field theory analogs of quasilocal lattice charges are therefore  necessary. As a concrete example, we find the complete set of charges needed in the sine-Gordon model, by using the fact that this field theory is recovered as the continuum limit of a spatially inhomogeneous version of the XXZ chain. The set of quasilocal charges of the lattice theory are shown to become a set local charges with fractional spin in the field theory.
\end{abstract}

\section{Introduction}
\label{sec:intro}

The last decade has been the stage of formidable progress in the study of many-body quantum systems out of equilibrium, motivated to a large extent by important advances in ultra-cold atomic experiments \cite{Kinoshita,Bloch,Giamarchi,Polkovnikov,cetal-12,gklk-12, tcfm-12,lgkr-13,mmk-13,fse-13,legr-15,adrl-14,ktlr-16} now able to realize and analyze almost perfectly isolated quantum systems. This has provided a way to directly observe the unitary time evolution following a quantum quench \cite{cc-06}, where a system is prepared in an arbitrary initial state and let evolve, and has shown the central role played by the existence of conservation laws in the out-of-equilibrium dynamics. In particular, in integrable systems such as spin chains or continuous models described by relativistic or non-relativistic quantum field theories \cite{muss1,fiorettomussardo,muss3,cardy1,cardy2,cardy3,doyonquasiloc,doyonquasiloc2,alvisespyros,milosz2}, the presence of a large number of non-trivial conservation laws imposes strong constraints on the time evolution. 
In this case there is now accumulated evidence that equilibration of local observables should occur at long times towards a generalized Gibbs ensemble (GGE) \cite{Rigol,rdyo-07,vr-16,ef-16,cef-11,fe-13,fe2-13,caza,bs-08,cdeo-08,scc-09,csc-13,kcc-14,sc-14,ga-15,pe-16,cem-16}, 
defined from an appropriate set of (quasi)local conserved charges. 

Over the recent years, a significant role has been played by the introduction of the so called quench-action approach \cite{CauxEssler,brockmann,CauxQA}. According to the latter, the post-quench steady-states of Bethe ansatz integrable systems can be characterized in terms of a single class of {\it representative eigenstates}, parametrized by a set of densities of generalized momenta (the so-called Bethe roots) and bound-states thereof (the so-called strings). A major task left by the quench-action approach is therefore to determine the set of densities from a given initial state. This has already been achieved in several experimentally-relevant cases \cite{brockmann,dwbc-14,pce-16,bucciantini-16,alca-16,bertiniShG,Bertini,PozsgayGGE1,Wouters}, and has confirmed the full ability of the quench-action, when manageable, to completely determine the steady-states properties following a quantum quench. Complementarily, this has also helped shaping the understanding of the GGE, namely which charges the latter should include in order to properly determine the post-quench representative densities.  
In this perspective, it has recently been observed in the context of quantum spin chains that a GGE built out of the traditionally known set of local charges does not completely reproduce the steady-state properties \cite{FagottiGGE,PozsgayGGE1,Wouters,PozsgayGGE2,PozsgayGGE3}, and should instead be complemented by newly discovered quasilocal charges \cite{prosen-14,imp-15,zmp-16,impz-16} whose construction follows from the seminal works of Prosen and collaborators \cite{prosen-11,prosen-13}. This has further been related to the fact that local charges do not provide any information about the densities of various strings, which is instead contained in quasilocal charges \cite{Ilievski-GGE,Ilievski-stringcharge,pvc-16,pvcr-16}.

In the present work we focus on the case of relativistic integrable quantum field theories (QFT), in particular on the sine-Gordon model. The latter is a prototypical example of a theory with  {\it non-diagonal scattering}, namely characterized by different species of particles labeled by an index $i$ (in sine-Gordon these are the soliton and antisoliton, as well as the possible bound states thereof, the so-called breathers) whose individual momenta are conserved but whose index $i$ may change through scattering processes. 
The thermodynamic Bethe Ansatz \cite{ZamoloTBA} construction of eigenstates in non-diagonal theories uses the strategy of replacing the difficult problem of non-diagonal scattering by one of a diagonal scattering theory with some additional auxilliary particles, the so-called {\it magnons}, which carry no momentum or energy but whose role is to specify the configuration of internal indices $i$ of the physical particles. Therefore, a given eigenstate is in this case specified not only by a set of densities of quasi-momenta for each type of physical particles, but also by the densities 
associated to each type of magnons and bound states thereof, the so-called {\it magnonic strings}. 
We point out that quantum quenches of sine-Gordon have already been studied using form factor expansions in \cite{demlerSG}, and using the quench-action method in \cite{Bertini}. In the latter case, the representative state contained no information about the magnons. This was however not a central problematic issue, since in \cite{Bertini} only neutral vertex operators were studied, which only have non vanishing matrix elements on states with equal number of solitons and antisolitons. Similar results for sine-Gordon were also obtained in \cite{marton}, using an alternative semiclassical approach, looking at the same neutral initial states (see also \cite{marton2} for an extension of this method to further results). 
The knowledge of the magnonic densites however becomes essential as soon as one is interested in measuring the post-quench properties of observables which are not neutral with respect to the internal exchange of the indices $i$. 

In most relativistic integrable quantum field theories, the construction of a set local conserved charges is well-known (see \cite{Giuseppebook} for a review): these charges $\{Q_s\}$ are characterized by an integer Lorentz spin $s$, and include in particular the momentum and energy operators with $s=1$. In the case of theories with non-diagonal scattering, these are, however, blind to the index content of the various particles and are therefore unable to fix the various densities of magnonic and strings determining the steady-state properties after a quantum quench. It was shown in \cite{milosz} that even in diagonal-scattering theories, this discrete set of integer-spin charges is not always sufficient to completely determine a state, but a continuous set of quasilocal conserved charges needs to be included in the GGE. The procedure to obtain these charges is reviewed in Section \ref{sec:GGESG}.

Our aim in the following is to make progress towards the construction of a complete GGE in theories with non-diagonal scattering by constructing a new set of conserved charges in relation with the diverse species of magnonic strings, in the prototypical case of the sine-Gordon model in its repulsive regime (where no breathers exist). The construction relies on the fact that this model can be studied as the continuum limit of a spatially inhomogeneous deformation of the Heisenberg spin chain, following the so-called ``light-cone" discretization \cite{DDV-MT, DDV-RSOS, DDV-NLIEexcited, SalResh}. We show that the definition of the lattice quasilocal charges of \cite{Ilievski-GGE,Ilievski-stringcharge} can be extended to the inhomogeneous case, and leads in the continuum limit to a new set of {\it local} conserved charges which, importantly, act non-trivially on the various configurations of magnons. More precisely, we construct in the sine-Gordon limit the action of our charges on the various eigenstates rather than an explicit operatorial expression. This implies in particular that a direct implementation of the GGE, which necessits evaluating the expectation values of conserved charges on arbitrary initial state, requires some further work.

There is an interesting interpretation of the Lorentz spin of our new conserved charges in field theory. As we have mentioned, the usual local conserved charges in integrable QFT have integer spin. In field theories that can be associated with some lattice model, it may be shown that the  set of integer-Lorentz-spin field theory charges arises from a set of {\it local} conserved charges on the lattice which have finite support on an integer number of lattice sites. It then seems naturally intriguing to find what is the spin of the charges in field theory which arise from quasilocal lattice charges, which have support on all lattice sites, though with an exponentially decaying norm. A striking feature of our new charges is that, contrarily to the aforementioned set of local ones $\{Q_s\}$ (which for sine-Gordon has $s=1,3,5,\ldots$), they have {\it fractional} Lorentz spins, which depend on the coupling. Some other instances of fractional spin charges may be found in the sine-Gordon litterature \cite{Bernard1, Bernard2, Russians}, however these were defined in an intrinsically non-local way. We will argue in the following that some relation might nevertheless exist between the various constructions.

The plan of the paper is the following. In Section \ref{sec:TBAnondiag} 
we review the thermodynamic Bethe ansatz (TBA) construction of eigenstates for field theories with non-diagonal scattering, with special attention brought to the sine-Gordon model. We thereby introduce the different species of magnons and the associated densities.  
In Section \ref{sec:lightcone}, we introduce the light-cone discretization of the sine-Gordon theory. In \ref{sec:charges} we construct the light-cone lattice quasilocal charges and their sine-Gordon limit. We compare our construction to that of the previously known fractional spin charges of \cite{Bernard1,Bernard2} and \cite{Russians}, which we review in more detail in Appendix \ref{app:othercharges}. 
In Section \ref{sec:GGESG}, we describe how our charges may be used to implement a complete GGE. In particular, we review how the scaling limit of these charges should be taken from the lattice, following the construction of \cite{milosz} for diagonal theories. 
The paper contains two more appendices. In Appendix \ref{sec:XXZ} we review some results about the Bethe ansatz solution of the XXZ spin chain, which is used extensively in our construction. In Appendix \ref{app:quasilocal}, we present a technical proof of the quasilocality of the constructed lattice operators.

\section{Quantum field theories with non-diagonal scattering}
\label{sec:TBAnondiag}

In this section we describe the thermodynamic Bethe ansatz \cite{ZamoloTBA} construction of eigenstates for integrable QFTs with non-diagonal scattering.  This will make clear what are the densities that need to be fixed in the QA approach, which need to be reproduced by the GGE. 

In an integrable field theory, due to the fact that all scattering is completely elastic, the momentum occupation modes $I(\theta)=Z^\dag(\theta)Z(\theta)$ are conserved, where $Z^\dag(\theta)$ and $Z(\theta)$ are particle creation and annihilation operators, respectively, and $\theta$ are the rapidities, related to the particle energy and momentum by $E=m\cosh\theta$, $p=m\sinh\theta$. Unlike the local charges $Q_s$ discussed in the previous section, we point out that these operators, $I(\theta)$ are highly nonlocal, and  not extensive quantities that should be included in the GGE.

 The one particle asymptotic particle states in a non-diagonal QFT can be written as 
\begin{eqnarray}
\vert \theta, i\rangle=Z_i^\dag(\theta)\vert 0\rangle,\nonumber
\end{eqnarray}
where $i$ is some index denoting different species of particles. Non-diagonal scattering means that while their rapidities are conserved, particles can change the value of this index in a scattering process. 
The two-particle S-matrix, $S_{ij}^{lk}(\theta)$ (which is a non-diagonal matrix in terms of the indices, hence the name: non-diagonal scattering), can be used to define the Fadeev-Zamolodchikov algebra for particle-creation and annihilation operators
\begin{eqnarray}
&Z_i^\dag(\theta_1)Z_j^\dag(\theta_2)=S_{ij}^{lk}(\theta_1-\theta_2)Z_k^\dag(\theta_2)Z_{l}^\dag(\theta_1),\nonumber\\
&Z_i(\theta_1)Z_j^\dag(\theta_2)=2\pi\delta_{ij}\delta(\theta_1-\theta_2)+S_{ij}^{lk}(\theta_2-\theta_1)Z_l^\dag(\theta_2)Z_k(\theta_1)
.\nonumber
\end{eqnarray}

A prominent example we will have in mind, and to which we will specify in most of this paper, is that of the sine-Gordon theory defined in terms of a real field $\varphi(x,t)$ by the action  \cite{ZamoloSG}
\be
\mathcal{A} = \int d^2 x \left(  
\frac{1}{2} (\partial_\nu \varphi)^2
- 2 m_0 \cos (\beta \varphi) \,,
 \right)  \,.
 \label{eq:SGaction}
 \nonumber
\ee
In this case the elementary particles are the soliton and antisoliton (so the particle index runs over $i=a,s$), classically thought as field configurations interpolating between two adjacent minima of the $\cos(\beta \varphi)$ potential. For some values of the coupling $\beta$, the interactions between solitons and antisolitons are attractive, and bound states can also be formed, which are the so called ``breathers". For most of this paper we will focus on the repulsive regime $\beta > \sqrt{4 \pi}$ where there are no bound states.
We also introduce the parameters $p$ and $\gamma$ as 
\be
\frac{\beta^2}{8 \pi} = \frac{p}{p+1}  = 1- \frac{\gamma}{\pi}\,,
\label{eq:parameterp}
\ee
that is  
\be 
\gamma = \frac{\pi}{p+1} \,,
\ee 
in terms of which the repulsive regime corresponds to $p > 1, \gamma < \frac{\pi}{2}$.

\subsection{Thermodynamic Bethe ansatz}

 We consider first an interacting diagonal theory with one species of particle. The scattering matrix then reduces to a scattering amplitude $S(\theta)$. 
In finite volume $L$, the momentum occupation modes $I(\theta)=Z^\dag(\theta)Z(\theta)$ are not independent. 
 When periodic boundary conditions are imposed, the coupling between modes is given by the $N$-particle Bethe ansatz quantization condition,
\begin{eqnarray}
e^{i L m\sinh\theta_n}\prod_{m\neq n}S(\theta_n-\theta_m)=\pm 1,\,\,\,\,n=1\dots,N,\label{betheone}
\end{eqnarray}
where the $\pm$ is chosen by the sign of $S(0)=\pm 1$. It is also convenient to re-express (\ref{betheone}) by taking the logarithm on both sides, which gives
\begin{eqnarray}
m L\sinh\theta_i+\sum_{j\neq i}\delta(\theta_i-\theta_j)=2\pi n_i,\label{bethetwo}
\nonumber
\end{eqnarray}
where $\delta(\theta)=-i\ln S(\theta)$ and the numbers $\{n_i\}$ are integers for $S(0)=+1$, and half integers for $S(0)=-1$.

In the thermodynamic limit, taken by letting $N,L\to\infty$ while $N/L$ is kept finite, the allowed particle rapidities that are solutions of (\ref{betheone}) are very close to each other, with the distance between two adjacent solutions being of order $(\theta_i-\theta_{i+1})\approx 1/m L$. It then becomes convenient to introduce the continuous density, $\rho(\theta)$, defined as the number of particles with rapidity between $\theta$ and $\theta+\Delta\theta$ divided by $L\Delta\theta$. The quantization condition (\ref{betheone}) becomes in the thermodynamic limit
\begin{eqnarray}
\frac{m}{2\pi}\sinh \theta_i+(\delta \star \rho)(\theta_i)=\frac{n_i}{L},\label{bethethree}
\end{eqnarray}
where $\star$ denotes the convolution 
\begin{eqnarray}
(f\star g)(\theta)=\int_{-\infty}^{\infty}\frac{d\theta^\prime}{2\pi}f(\theta-\theta^\prime)g(\theta^\prime)
\,
.\nonumber
\end{eqnarray}
Of the allowed momentum modes that are solutions of (\ref{bethethree}) those that are occupied are called roots, and those not excited are called holes, which can also be described by a hole-density distribution, $\rho^{(h)}(\theta)$ in the thermodynamic limit. The root and hole densities are related by the condition
\begin{eqnarray}
\rho(\theta)+\rho^{(h)}(\theta)=\frac{1}{2\pi}m\cosh \theta+(\varphi \star \rho)(\theta),\nonumber
\end{eqnarray}
where 
\begin{eqnarray}
\varphi(\theta)=\frac{d}{d\theta}\delta(\theta).\nonumber
\end{eqnarray}

Let us now step up to the case of non-diagonal theories.  
We first need to find what is the quantization condition for an $N$-particle state in a finite volume $L$. We introduce the $N$-particle monodromy matrix, which represents the scattering of one particle of rapidity $\lambda$ with the $N$ particles around the finite volume, with index value $a$ before scattering, and $b$ after scattering with all the particles, 
\begin{eqnarray}
\mathcal{M}(\lambda\vert\{\theta_k\})_{a,\{i_k\}}^{b,\{j_k\}}=S_{a\, i_1}^{c_1 j_1}(\lambda-\theta_1)S_{c_1 i_2}^{c_2 j_2}(\lambda-\theta_2)\dots S_{c_{N-1} i_N}^{b\, j_N}(\lambda-\theta_N).\nonumber
\end{eqnarray}
For sine-Gordon \cite{FeherTakacs,Takacs-NLIE}, this is a $2\times2$ matrix in the indices $a,b$ which can be written as 
\begin{eqnarray}
\mathcal{M}(\lambda\vert\{\theta_k\})_{\{i_k\}}^{\{j_k\}}=
\left(\begin{array}{cc}
A(\lambda\vert\{\theta_k\})_{\{i_k\}}^{\{j_k\}}&B(\lambda\vert\{\theta_k\})_{\{i_k\}}^{\{j_k\}}\\
C(\lambda\vert\{\theta_k\})_{\{i_k\}}^{\{j_k\}}&D(\lambda\vert\{\theta_k\})_{\{i_k\}}^{\{j_k\}}\end{array}\right).
\nonumber
\end{eqnarray}
One can also define the transfer matrix as the trace of the monodromy matrix (trace over the indices $a,b$)
\begin{eqnarray}
\mathcal{T}(\lambda\vert\{\theta_k\})_{\{i_k\}}^{\{j_k\}} =\mathcal{M}(\lambda\vert\{\theta_k\})_{a\{i_k\}}^{a\{j_k\}}=A(\lambda\vert\{\theta_k\})_{\{i_k\}}^{\{j_k\}}+D(\lambda\vert\{\theta_k\})_{\{i_k\}}^{\{j_k\}}.\nonumber
\end{eqnarray}
With this definition, the Bethe quantization condition on some wave $N$-particle wave function, $\Psi(\{\theta_k\})_{\{i_k\}}$, analogous to (\ref{betheone}) can be written simply as
\begin{eqnarray}
e^{i L m\sinh\theta_j}\mathcal{T}(\theta_j\vert\{\theta_k\})_{\{i_k\}}^{\{j_k\}}\Psi(\{\theta_k\})_{\{j_k\}}=\Psi(\{\theta\})_{\{i_k\}}\label{betheonenondiagonal}
\end{eqnarray}

Transfer matrices with different parameters can be shown to commute with each other,
\begin{eqnarray}
\left[\mathcal{T}(\lambda\vert\{\theta_k\})_{\{i_k\}}^{\{j_k\}},\mathcal{T}(\mu\vert\{\theta_k\})_{\{i_k\}}^{\{j_k\}}\right]=0,
\nonumber
\end{eqnarray}
which means they can be simultaneously diagonalized. The condition (\ref{betheonenondiagonal}) can be solved by finding the eigenfunctions  of the transfer matrix. In sine-Gordon  \cite{FeherTakacs,Takacs-NLIE} these can be found starting from a ``reference state", $\Psi_{0}(\{\theta_k\})_{\{i_k=s\}}$ for which all the particles are solitons, with eigenvalue equation
\begin{eqnarray}
\mathcal{T}(\lambda\vert\{\theta_k\})_{\{i_k\}}^{\{j_k\}}\Psi_{0}(\{\theta_k\})_{\{j_k\}}&=&\left[A(\lambda\vert\{\theta_k\})_{\{i_k\}}^{\{j_k\}}+D(\lambda\vert\{\theta_k\})_{\{i_k\}}^{\{j_k\}}\right]\Psi_{0}(\{\theta_k\})_{\{j_k\}}.\nonumber\\
&=&\!\!\!\! \Lambda_0(\lambda\vert\{\theta_k\})\Psi_{0}(\{\theta_k\})_{\{i_k\}}.
\nonumber
\end{eqnarray}

One can find eigenstates of the transfer matrix with $Q$ antisolitons and $N-Q$ solitons by acting on the reference state as
\begin{eqnarray}
\Psi_Q(\{\lambda_l\}\vert \{\theta_k\})_{\{i_k\}}=\left[\prod_{l=1}^Q B(\lambda_l+\frac{i\pi}{2}\vert\{\theta_k\})^{\{j_k\}}_{\{i_k\}}\right]\Psi_0(\{\theta_k\})_{\{i_k\}} \,, 
\nonumber
\end{eqnarray}
where the parameters $\{ \lambda_l\}$ satisfy
\begin{eqnarray}
\mathcal{T}(\lambda \vert\{\theta_k\})_{\{i_k\}}^{\{j_k\}}\,\Psi_Q(\{\lambda_l\}\vert \{\theta_k\})_{\{j_k\}}=\Lambda(\lambda,\{\lambda_l\}\vert\{\theta_k\})\,\Psi_Q(\{\lambda_l\}\vert \{\theta_k\})_{\{i_k\}}.
\nonumber
\end{eqnarray}
The set of parameters $\{\lambda_l\}$ can be thought of as a set of rapidities of some auxiliary particles. These auxiliary particles act as a wave that moves with rapidity $\lambda_l$ changing the values of the indices $i_k$ as it passes. These waves carry no energy or momentum, and are typically called magnons.

The quantization condition (\ref{betheonenondiagonal}) can now be rewritten as a quantization condition of a theory with diagonal scattering of real particles and magnons. For a state with $N$ real particles and $Q$ magnons, the quantization conditions for real particles are
\begin{eqnarray}
e^{i L m\sinh\theta_j}\Lambda(\theta_j,\{\lambda_l\}\vert\{\theta_k\})=1, \,\,j=1,\dots,N\label{bethetwonondiagonal}
\end{eqnarray}
with an auxiliary quantization condition for the magnons
\begin{eqnarray}
\Lambda(\lambda_j,\{\lambda_k\}\vert\{\theta_k\})=1,\,\,j=1,\dots,Q\label{bethethreenondiagonal} \,. 
\end{eqnarray}
For sine-Gordon, equations (\ref{bethetwonondiagonal},\ref{bethethreenondiagonal}) read explicitely \cite{FeherTakacs} (note the shift of the magnonic rapidities $\lambda_j$ by $i \pi/2$ with respect to those of \cite{FeherTakacs}) 
\bea
\mathrm{e}^{ i m L \sinh \theta_j} 
&=&  
\prod_{k(\neq j)}^N S_0 \left(\theta_j- \theta_k \right)
\prod_{k=1}^{Q} 
\frac{\sinh  \frac{1}{p}\left(\theta_j - \lambda_k + i \pi/2\right)  }{\sinh  \frac{1}{p}\left(\theta_j - \lambda_k - i \pi/2\right)} 
\label{eq:SGTBA1}
\\
\prod_{k=1}^N 
\frac{\sinh  \frac{1}{p}\left(\lambda_j - \theta_k + i \pi/2\right)  }{\sinh  \frac{1}{p}\left(\lambda_j - \theta_k - i \pi/2\right)} 
&=&
\prod_{k (\neq j)}^Q 
\frac{\sinh \frac{1}{p}\left(\lambda_j - \lambda_k + i \pi \right)  }{\sinh  \frac{1}{p}\left(\lambda_j - \lambda_k - i \pi \right)} 
 \,,
\label{eq:SGTBA2}
\eea 
where the scattering amplitude $S_0(\theta)$ is given by 
\be
S_0(\theta) = \mathrm{exp} \left(
i \int_{0}^{\infty} \frac{dt}{t} \frac{\sinh\left( \frac{p-1}{2} t\right)}{\sinh \frac{p}{2}t} \frac{\sin t\theta /\pi}{\cos t/2}
\right)  \,.
\label{eq:SGSmatrix}
\ee

From this point onwards, the quantization conditions (\ref{bethetwonondiagonal}) and (\ref{bethethreenondiagonal}) are identical to a problem of diagonal scattering with two species of particles, one of which is auxiliary and carries no energy.
As we will now see, in sine-Gordon (but similar comments also hold true in other non-diagonal scattering theories), the aforementioned magnons can further form bound states which effectively appear as new species of particles. The classification of possible bound states in fact depends in an intricate way on the coupling $\beta$, and will be reviewed in some detail in the next paragraph. 

Before closing this section, we however would like to comment on the attractive regime ($\beta \leq \sqrt{4\pi}$, $p \leq 1$). There, as mentioned earlier, the solitons and antisolitons can themselves form bound states, the so-called {\it breathers}, the different possible kinds of which also depends on the value of the parameter $\beta$. The scattering matrices between breathers and solitons, breathers and magnons, and between two different breathers is determined self-consistently, and the Bethe equations (\ref{eq:SGTBA1},\ref{eq:SGTBA2}) may be recast into a different, more complicated form. Though we will not study this regime in detail in the present work, we will describe some aspects of how the charges we are going to construct are expected to behave in the attractive case.

\subsection{Bethe ansatz for strings in sine-Gordon at rational values of the coupling}
\label{sec:stringsSG}

As mentioned above, the magnonic Bethe roots of (\ref{eq:SGTBA1},\ref{eq:SGTBA2}) may form bound states whose classification depends on the parameter $\beta$. 
The structure of these bound states may in fact be deduced from the observation that the magnonic Bethe equations (\ref{eq:SGTBA2}) coincide with the Bethe equations (\ref{eq:bethe_eq}) presented in Appendix \ref{sec:XXZ} for the XXZ chain in its gapless phase, provided the parameter $\tilde{p}$ of Appendix \ref{sec:XXZ} is taken as
\be
\tilde{p} = p-1 \,,
\ee
and that the inhomogeneous parameters $\{v_j \}$ are chosen as 
\be
v_j =  \frac{1}{p}\theta_i \,.
\ee
The solutions of the XXZ Bethe ansatz equations have been extensively studied in the past litterature \cite{Takahashi}, and some results are reviewed in Appendix \ref{sec:XXZ}. In particular, it is observed that the roots $\{\lambda_k\}$ assemble into regular patterns called {\it strings}. In the present conventions, one defines an (even parity) $m$-string as a set of magnonic roots of the form
\be
\lambda_{k}^{\nu,(m)} = \lambda_{k}^{(m)} + i \pi \left(\nu - \frac{m+1}{2} \right) + \delta^{\nu, (m)}_k  \,, \qquad \qquad \nu = 1,\ldots  m \,,
\nonumber
\ee
where $\lambda^{(m)}_{k}$ is a real number called the string center, and the numbers $\delta^{\nu,(m)}_{k}$ are deviations from a perfect string which vanish exponentially with the system size and are therefore neglected in the so-called {\it string hypothesis} \cite{Takahashi}\footnote{While in \cite{Takahashi} the string hypothesis is only stated for the homogeneous case, it is easy to see that the arguments which lead to it still apply in the case of imaginary inhomogeneous parameters, namely, in the present notations, for arbitrary {\it real} values of the parameters $v_j$.}. In addition one may also encounter strings of odd parity, the so-called $(m-)$-strings, whose center $ \lambda^{(m-)}_{k}$ is shifted by $i p \frac{\pi}{2}$. 
The set of allowed string species depend in an important manner on the parameter $\tilde{p}$. 
Through the gapless phase (namely, for real values of the parameter $\tilde{p}$), it turns out to be convenient to restrict the study to the dense set of rational values of $\tilde{p}$, for which the set of allowed string is well-determined \cite{Takahashi}.

To keep the presentation simple, we will here consider in detail the case where $p= \tilde{p}+1$ is an integer $\geq 1$. Transposing the results of Appendix \ref{sec:XXZ}, the set of allowed strings is in this case
\bea
\mbox{ even parity : } m&=&1,2,\ldots p-1 \,, \nonumber \\
\mbox{odd parity : }& &(1-) \,. \nonumber 
\eea
For other rational values of $p$ the string content is generally different, but in spirit everything that will be described in the following may be readily adapted. 

In the thermodynamic limit, $N, L \to \infty$, $N/L$ finite,  the Bethe quantization equations (\ref{eq:SGTBA2}) can be recast into a linear form for the densities of various string centers or holes thereof, $\rho_{m}, \rho_{m}^h$.
These have the form 
\be
\rho_m(\lambda) + \rho_m^h(\lambda) =  \frac{1}{L} \sum_{j=1}^N {a}_m(\lambda- \theta_j) - \left(\sum_{n} {a}_{m,n} \star \rho_n\right)(\lambda) \qquad \qquad (m= 1,\ldots, p-1,(1-)) \,,
\label{eq:BetheTakahashitilde} 
\ee
where the kernels are those of Appendix \ref{sec:XXZ} (equation (\ref{eq:BAEkernels})), with the replacement $\tilde{p} = p-1$.
One difference between (\ref{eq:BetheTakahashitilde}) and their XXZ equivalent (\ref{eq:BetheTakahashi}) is that in the present case the densities of rapidities are defined with respect to the physical length $L$, rather than the number $N$ of magnons. 

The equations (\ref{eq:BetheTakahashitilde}) can be recast into a very useful form by introducing the inverse kernel. 
Namely, convoluting (\ref{eq:BetheTakahashi}) with the inverse kernel \cite{Takahashi,ZamoloTBA} 
\be 
C_{m,n}(\lambda) = \left(a +  \mathbf{\delta} \right)_{m,n}^{-1} (\lambda)
= \delta_{m,n}\mathbf{\delta}(\lambda) - s(\lambda) I_{m,n} \,,
\nonumber
\ee 
where
\be
s(\lambda) = \frac{1}{\cosh (\pi \lambda)}   \,,
\label{eq:sdef}
\ee
 $\mathbf{ \delta}$ is the Dirac delta function in $\lambda$ space, while $I$ is a matrix specified by the following non-zero entries 
\bea
I_{m,n} &=&  \delta_{m,n+1} + \delta_{m,n-1}  \qquad\qquad \mbox{($m=1, \ldots p-3$, $n=1, \ldots p-2$)} \nonumber \\
I_{p-2,n} &=&  \delta_{p-3,n} + \delta_{p-1,n}+\delta_{(1-),n}   \nonumber \\
I_{p,n} &=&  I_{(1-),n} = \delta_{p-2,n} \,, 
\eea
one obtains the equations 
\be
\rho_m^h + \rho_m = \delta_{m,1}~s \star \rho + I_{m,n} s \star \rho_n^h  \qquad \qquad (m= 1,\ldots, p-1,(1-)) \,.
\label{eq:Dynkin}
\ee
where $\rho(\theta)$ is the density of physical particles.

These equations can be conveniently encoded in a Dynkin diagram \cite{DynkinTBA, Balog,Gromov}, with one node for each string species, and of which $I$ is the incidence matrix. We represent it on the bottom-right panel of figure \ref{fig:strings}.  
In this representation each node stands for a string type, a link joining two nodes $m$ and $n$ stands for a term in $\rho_m^h$ entering the equation (\ref{eq:Dynkin}) for $\rho_n$, while the red node indicates that the density $\rho(\theta)$ of physical particles acts as a source in the equation for $\rho_1$.

In summary, the eigenstates of the sine-Gordon model in its repulsive regime can be described in the thermodynamic limit by a set of densities $\rho_m(\lambda), \rho_{m}^h(\lambda)$ for each allowed type of magnonic string, in addition to the densities $\rho(\theta), \rho^h(\theta)$ for the physical particles and holes thereof. 
The aim of the complete GGE we want to build is therefore to be able to fix all these densities completely. To proceed in this direction, we will use a discretization of the sine-Gordon model in terms of an inhomogeneous lattice model.

\section{The light cone discretization of the sine-Gordon model}
\label{sec:lightcone}

\subsection{The light cone lattice}

Field theories are customarly regularized by discretizing the space and time directions separately, and encoding the discrete time evolution in the so-called row-to-row transfer matrix. 
We will here instead use a different scheme, namely the so-called light-cone discretization, introduced by Destri and de Vega in \cite{DDV-MT} and further used in \cite{DDV-RSOS, DDV-NLIEexcited, SalResh}, and which we review in some detail here. 

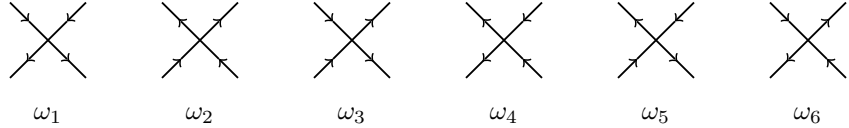
\begin{figure}
\begin{center}
\begin{tikzpicture}[scale=1]
\begin{scope}[shift={(-4,0)}]
\draw[-<, line width=0.7] (-0.5,-0.5) -- (-0.22,-0.22);  
\draw[line width=0.7] (-0.25,-0.25) -- (0.25,0.25);
\draw[<-, line width=0.7] (0.25,0.25) -- (0.5,0.5);
\draw[-<, line width=0.7] (0.5,-0.5) -- (0.22,-0.22);
\draw[line width=0.7] (0.25,-0.25) -- (-0.25,0.25);
\draw[<-, line width=0.7] (-0.25,0.25) -- (-0.5,0.5);
\node at (0,-1) {$\omega_1$};
\end{scope}
\begin{scope}[shift={(-2,0)}]
\draw[->, line width=0.7] (-0.5,-0.5) -- (-0.25,-0.25); 
\draw[line width=0.7] (-0.25,-0.25) -- (0.25,0.25);
\draw[>-, line width=0.7] (0.22,0.22) -- (0.5,0.5);
\draw[->, line width=0.7] (0.5,-0.5) -- (0.25,-0.25);
\draw[line width=0.7] (0.25,-0.25) -- (-0.25,0.25);
\draw[>-, line width=0.7] (-0.22,0.22) -- (-0.5,0.5);
\node at (0,-1) {$\omega_2$};
\end{scope}
\begin{scope}[shift={(-0,0)}]
\draw[->, line width=0.7] (-0.5,-0.5) -- (-0.25,-0.25);
\draw[line width=0.7] (-0.25,-0.25) -- (0.25,0.25);
\draw[>-, line width=0.7] (0.22,0.22) -- (0.5,0.5);
\draw[-<, line width=0.7] (0.5,-0.5) -- (0.22,-0.22);
\draw[line width=0.7] (0.25,-0.25) -- (-0.25,0.25);
\draw[<-, line width=0.7] (-0.25,0.25) -- (-0.5,0.5);
\node at (0,-1) {$\omega_3$};
\end{scope}
\begin{scope}[shift={(2,0)}]
\draw[-<, line width=0.7] (-0.5,-0.5) -- (-0.22,-0.22);
\draw[line width=0.7] (-0.25,-0.25) -- (0.25,0.25);
\draw[<-, line width=0.7] (0.25,0.25) -- (0.5,0.5);
\draw[->, line width=0.7] (0.5,-0.5) -- (0.25,-0.25);
\draw[line width=0.7] (0.25,-0.25) -- (-0.25,0.25);
\draw[>-, line width=0.7] (-0.22,0.22) -- (-0.5,0.5);
\node at (0,-1) {$\omega_4$};
\end{scope}
\begin{scope}[shift={(4,0)}]
\draw[->, line width=0.7] (-0.5,-0.5) -- (-0.25,-0.25); 
\draw[line width=0.7] (-0.25,-0.25) -- (0.25,0.25);
\draw[<-, line width=0.7] (0.25,0.25) -- (0.5,0.5);
\draw[-<, line width=0.7] (0.5,-0.5) -- (0.22,-0.22);
\draw[line width=0.7] (0.25,-0.25) -- (-0.25,0.25);
\draw[>-, line width=0.7] (-0.22,0.22) -- (-0.5,0.5);
\node at (0,-1) {$\omega_5$};
\end{scope}
\begin{scope}[shift={(6,0)}]
\draw[-<, line width=0.7] (-0.5,-0.5) -- (-0.22,-0.22);  
\draw[line width=0.7] (-0.25,-0.25) -- (0.25,0.25);
\draw[>-, line width=0.7] (0.22,0.22) -- (0.5,0.5);
\draw[->, line width=0.7] (0.5,-0.5) -- (0.25,-0.25);
\draw[line width=0.7] (0.25,-0.25) -- (-0.25,0.25);
\draw[<-, line width=0.7] (-0.25,0.25) -- (-0.5,0.5);
\node at (0,-1) {$\omega_6$};
\end{scope}
\end{tikzpicture} 
\end{center}
\caption{Vertices and weights of the six-vertex model. 
}
\label{fig:sixv}
\end{figure}

The light-cone discretization of the sine-Gordon model was initially formulated in terms of the equivalent massive Thirring model, which is a relativistic field theory for a massive fermion with quartic interactions. The two-dimensional Minkowski space-time is discretized as a diagonal square lattice of spacing $\delta$, associated with the light-cones coordinates $x_\pm = x \pm t$. The diagonal links are the possible world lines for left- or right-moving fermions, and are given an upwards/downwards orientation according to whether they are occupied by a fermion or empty. 
Interactions between fermions, as well as their mass, take place at the sites of the light-cone lattice, and are encoded in the respective weights associated with each of the six possible vertices represented on figure \ref{fig:sixv}. 
The corresponding weights should be chosen as those of the integrable six-vertex model, namely 
\be
\omega_1 = \omega_2 = 1 \,,\qquad 
\omega_3 = \omega_4 = \frac{\sin(i \Lambda - \gamma)}{\sin i \Lambda} \,,\qquad 
\omega_5 = \omega_6 = \frac{1}{\sin i \Lambda} \,, \qquad 
\ee
where $\Lambda$ is a real number which needs to be properly scaled when taking the continuum limit $\delta \to 0$ (see below), and the parameter $\gamma$ is that of equation (\ref{eq:parameterp}).

In this framework, the discrete time evolution of the sine-Gordon model is encoded by the so-called diagonal transfer matrix of the six-vertex model. The latter can however be reformulated through the row-to-row transfer matrix of an inhomogeneous six-vertex model, which we write as  
\be
T(u) = \mathrm{tr} \left(  R_{0,N}(u+  i \Lambda/2) R_{0,N-1}(u-   i \Lambda/2) \ldots R_{0,1} (u-   i \Lambda/2) \right)  \,. 
\label{eq:TXXZstaggered}
\nonumber
\ee
The transfer matrix (\ref{eq:TXXZstaggered}) acts on a row of $N$ (even) spins 1/2 (edges of the six-vertex model) with periodic boundary conditions, and is written as a trace over an auxilliary spin 1/2 of a product of $R$ matrices, encoding the weights of the six-vertex model as 
\be
R(u) = \frac{1}{\sin(\gamma-u)} 
\left( 
\begin{array}{cccc}
\sin(\gamma-u) & 0 & 0 & 0 \\
0 & {\sin u} & 1 & 0 \\
0 & 1 & {\sin u} & 0 \\
0 & 0 & 0 & \sin(\gamma-u) \\
\end{array}
\right)  \,.
\label{Rmatrix}
\ee

In the present work we define the momentum and Hamiltonian respectively as
\bea
P &=& \delta^{-1}  \left. \frac{\mathrm{d}}{\mathrm{d}u}  \log \left( T\left( i \frac{\Lambda}{2}+ i u\right) T\left( -i \frac{\Lambda}{2} - i u\right)\right)\right|_{u=0}
\label{eq:Pdef}
\\
H &=& \delta^{-1}  \left. \frac{\mathrm{d}}{\mathrm{d}u}  \log \left( T\left( i \frac{\Lambda}{2}+ i u\right) T\left( -i \frac{\Lambda}{2} - i u\right)^{-1} \right) \right|_{u=0}
\label{eq:Hdef} \,,
\eea
which differs by an order of derivation with respect to the usual conventions in the light-cone litterature \footnote{
In \cite{DDV-MT, DDV-RSOS, DDV-NLIEexcited, Takacs-NLIE} one defines instead 
\bea
P_{\rm light~cone} &=& \delta^{-1} i \log \left( T\left( i \frac{\Lambda}{2}\right) T\left( -i \frac{\Lambda}{2}\right) \right)  \label{eq:Plc}
\\
H_{\rm light~cone} &=& \delta^{-1}  i \log \left( T\left( i \frac{\Lambda}{2}\right) T\left( -i \frac{\Lambda}{2}\right)^{-1}\right) 
\,.
\label{eq:Hlc}
\eea  
}
and follows rather the conventions of \cite{SalResh}. This has the advantage of making $H, P$ local quantities on the lattice. As argued in \cite{SalResh} the two definitions are equivalent for the study of the scaling limit, however only the latter holds correct in a perturbative analysis. 

In the following we will show how the Hamiltonian (\ref{eq:Hdef}) may be diagonalized from Bethe ansatz, and how the correspondence with the sine-Gordon Bethe ansatz constructed in section \ref{sec:TBAnondiag} is recovered in the appropriate scaling limit.

\subsection{Bethe ansatz and scaling limit}
\label{sec:lightconescaling}

The transfer matrices (\ref{eq:TXXZstaggered}) form a mutually commuting family, and can be simultaneously diagonalized through Bethe ansatz, as is described in Appendix \ref{sec:XXZ}. 
In the present case, the parameter $\tilde{p}$ of Appendix \ref{sec:XXZ} has to be taken such that 
$
\gamma = \frac{\pi}{\tilde{p}+1} \,,
$
so it coincides precisely with the sine-Gordon parameter $p$ of equation (\ref{eq:parameterp}), 
and the staggered parameters $\{ u_j \equiv i v_j \}$ are here chosen as \\
 \be 
 v_{2i} = i \Lambda/2  \,, \qquad   v_{2i+1} = - i \Lambda/2  \,.
 \ee 
Here and in the following it will be convenient to introduce the rescaled staggering parameter
\be
\Theta = \frac{\pi}{\gamma} \Lambda = (p+1) \Lambda  \,.
\label{eq:tiltedLambda}
\ee
 
 The common eigenstates of the transfer matrices $T(u)$ (\ref{eq:TXXZstaggered}) are then parametrized by a set of quasimomenta $\{\alpha_k\}$, solution of the Bethe equations 
\be
\hspace{-0.5cm}
\left(
\frac{\sinh \left(\frac{1}{{p}+1} (\alpha_k - \Theta/2 + i \pi /2)  \right)}{\sinh \left(\frac{1}{{p}+1} (\alpha_k - \Theta/2 - i \pi /2) \right)} 
\right)^{N/2}
\left(
\frac{\sinh \left(\frac{1}{{p}+1} (\alpha_k + \Theta/2 + i \pi /2)  \right)}{\sinh \left(\frac{1}{{p}+1} (\alpha_k + \Theta/2- i \pi /2) \right)} 
\right)^{N/2}
= 
\prod_{l (\neq k)}
\frac{\sinh \left( \frac{1}{{p}+1}(\alpha_k - \alpha_l + i \pi ) \right)}{\sinh \left( \frac{1}{{p}+1} (\alpha_k - \alpha_l - i \pi ) \right) } \,. 
\label{eq:BAEstag}
\ee
The ground state is described by two Fermi seas of real roots $\alpha_i$, centered respectively around $\pm \Theta/2$. In the sine-Gordon / massive Thirring language this corresponds to the vacuum, while the massive excitations (namely, the sine-Gordon solitons/ massive Thirring fermions) are identified with holes of finite rapidities. Holes with rapidities $\sim \pm \Theta /2$ remain instead massless, and correspond to the excitations of an integrable QFT describing the RG flow between two conformal field theories \cite{SalResh}. In the scaling limit the massive and massless theories decouple, and we will be only interested in the former. The mass of the solitons/antisolitons is found to be \cite{DDV-MT,SalResh}
\be 
M \propto \delta^{-1} \mathrm{e}^{-\frac{\pi}{\gamma}\frac{\Lambda}{2}} \,,
\label{eq:physicalmass}
\ee
 while the bare mass can be found from the comparison of the Bethe equations (\ref{eq:BAEstag}) with those of the massive Thirring model \cite{BT,KorepinSG},
\be 
m_0 = 4 \sin\gamma \delta^{-1} \mathrm{e}^{-\Lambda} \,.
\nonumber
\ee

The {\it scaling limit}, yielding the sine-Gordon model on a circle of circumference $L$, is defined as
\bea
 N &\to& \infty   \,, \qquad \delta \to 0    \,, \qquad   \Lambda \to \infty    
 \nonumber
 \\ 
 L&=&  \delta N {\rm ~ fixed}  \,, \qquad
 \nonumber
 \\  
  M &\propto& \delta^{-1} \mathrm{e}^{-\frac{\pi}{\gamma}\frac{\Lambda}{2}} {\rm ~~ fixed} \,,
  \label{eq:scalinglimit} 
\eea
and must not be confused with the {\it thermodynamic limit} (or infinite  volume limit) which corresponds, once in the field theoretical setup (namely once (\ref{eq:scalinglimit}) taken) to sending $L \to \infty$ with a finite density of particles. 
We will now describe how, in the scaling limit, the Bethe ansatz equations (\ref{eq:BAEstag}) can be recast in a form which precisely reproduces the Bethe ansatz equations written in section \ref{sec:TBAnondiag} for the sine-Gordon model. This requires in particular to recast these equations in a form which has a well-defined limit as $N \to \infty$. 

For this sake, we follow the procedure described in \cite{Babelon} (for the periodic homogeneous XXZ model), \cite{DDV-RSOS} (for the staggered XXZ model with open boundary conditions), which consists in rewriting the Bethe ansatz equations (\ref{eq:BAEstag}) in terms of a finite set of excitations `on top' of the vacuum Fermi sea. The calculations presented below hold true throughout the repulsive regime $p>1$, while in the attractive regime several changes occur on which we will briefly comment shortly. 
In the repulsive regime, the various excitations are classified as follows : 
\begin{itemize}
\item  holes in the Fermi sea, with rapidities denoted as $\{ \theta_j \}$
\item complex conjugate pairs, among which $2$-strings ($|\mathrm{Im}|=\frac{\gamma}{2}$) and wide pairs ($|\mathrm{Im}|>\frac{\gamma}{2}$)
\item quartets of the form $\alpha +i\tau, \alpha -i\tau,  \alpha+i(\tau-\pi), \alpha-i(\tau-\pi)$, with $\alpha, \tau \in \mathbb{R}$ and $0 < |\tau| < \frac{\pi}{2}$.   
\end{itemize}
For each complex pair one then introduces a set of parameters $\lambda$: a 2-string is associated with a single parameter $\lambda$ which is the corresponding real part, while a wide pair $\alpha,\bar{\alpha}$ (with $\mathrm{Im}(\alpha) >0$) is associated with the complex pair $\lambda, \bar{\lambda} = \alpha-i\frac{\pi}{2}, \bar{\alpha}+i\frac{\pi}{2}$. Similarly, for each quartet of the form given above one introduces the parameters $\lambda, \bar{\lambda} = \alpha+i(\tau - \frac{\pi}{2}),  \alpha-i(\tau - \frac{\pi}{2})$. The set of parameters $\{\lambda_k\}_{k=1,\ldots Q}$ associated with the various complex excitations, together with the holes rapidities $\{\theta_j\}_{j=1,\ldots N_h}$, are shown to satisfy the higher-level equations
\bea
\mathrm{e}^{ i m L \sinh \theta_j}
&=&  
\prod_{k(\neq j)}^{N_h} S_0 \left(\theta_j - \theta_k \right)
\prod_{k=1}^{Q} 
\frac{\sinh \frac{1}{p} \left( \theta_j-\lambda_k + i\frac{\pi}{2}  \right)}{\sinh \frac{1}{p} \left( \theta_j-\lambda_k - i\frac{\pi}{2}  \right)} 
\label{eq:TBA1}
\\
\prod_{k=1}^{N_h}
\frac{\sinh \frac{1}{p} \left( \lambda_j-\theta_k + i\frac{\pi}{2}  \right)}{\sinh \frac{1}{p} \left( \lambda_j-\theta_k - i\frac{\pi}{2}  \right)} 
&=&  
\prod_{k (\neq j)}^{Q}
\frac{\sinh \frac{1}{p} \left( \lambda_j-\lambda_k + i\pi \right)}{\sinh \frac{1}{p} \left( \lambda_j-\lambda_k - i \pi  \right)}  \,,
\label{eq:TBA2}
\eea 
where $S_0$ is precisely the scattering amplitude (\ref{eq:SGSmatrix}) between sine-Gordon (anti)solitons. Equations (\ref{eq:TBA1}, \ref{eq:TBA2}) now have a finite $N\to \infty$ limit, and coincide with the sine-Gordon Bethe ansatz equations (\ref{eq:SGTBA1},\ref{eq:SGTBA2}). Note however that the number $N$ of physical particles in the latter has nothing to do with the number $N$ used in this section for denoting the number of lattice sites, instead it corresponds here to the number $N_h$ of hole like excitations.  

We now briefly go back to the question of the attractive regime. There, a similar approach can be used. However, as is well explained in \cite{DDV-NLIEexcited}, certain configurations of complex roots lie in different analyticity domains and the resulting higher-level Bethe equations are accordingly modified. This mechanism is in fact the one which leads to ``autonomous'' complex configurations, namely whose existence is independent from that of the holes and which in the sine-Gordon interpretation correspond to the breathers. 

\begin{figure}
\begin{center}
\begin{tikzpicture}[scale=1]
\begin{scope}[shift={(-4,2)}]
\node[rectangle,black, draw] at (-2.,1.25) {$\alpha_k$\,, \textcolor{red}{$\theta_i$}};
\draw (-3.5,0) -- (3.5,0);
\draw (0,-1.5) -- (0,1.5);
\draw[dashed] (-3.,0.25) node[left] {$\pi \over 2$}  -- (3,0.25);
\draw[dashed] (-3.,-0.25) node[left] {$-{\pi \over 2}$}  -- (3,-0.25);
\foreach \x in {-2,-1.75,...,2}
{ 
\fill[gray] (\x,0) circle (0.08);
}
\foreach \x in {-0.75,-0.25,0.5}
{ 
\fill[red] (\x,0) circle (0.1);
}
\foreach \y in {-0.75,-0.25,...,1}
{ 
\fill[black] (-1.1,\y) circle (0.08);
}
\foreach \y in {-1,-0.5,...,1}
{ 
\fill[black] (2.5,\y) circle (0.08);
}
\foreach \y in {-0.25,0.25}
{ 
\fill[black] (0.15,\y) circle (0.08);
}
\end{scope}
\begin{scope}[shift={(-4,-2)}]
\node[rectangle,black, draw] at (-2.,1.25) {$\lambda_k$\,, \textcolor{red}{$\theta_i$}};
\draw (-3.5,0) -- (3.5,0);
\draw (0,-1.5) -- (0,1.5);
\draw[dashed] (-3.,0.25) node[left] {$\pi \over 2$}  -- (3,0.25);
\draw[dashed] (-3.,-0.25) node[left] {$-{\pi \over 2}$}  -- (3,-0.25);
\foreach \x in {-0.75,-0.25,0.5}
{ 
\fill[red] (\x,0) circle (0.1);
}
\foreach \y in {-0.5,0,...,0.5}
{ 
\fill[black] (-1.1,\y) circle (0.08);
}
\foreach \y in {-0.75,-0.25,...,0.75}
{ 
\fill[black] (2.5,\y) circle (0.08);
}
\foreach \y in {0}
{ 
\fill[black] (0.15,\y) circle (0.08);
}
\end{scope}
\begin{scope}[shift={(2,2)}]
\fill[red] (0,0) circle (0.25);
\draw (0.25,0) -- (0.75,0);
\draw (1,0) circle (0.25);
\draw[dashed] (1.25,0) -- (2.75,0);
\draw (3,0) circle (0.25);
\draw (3.25,0) -- (3.75,0);
\draw (4,0) -- (4.75,0.8);
\draw (4,0) -- (4.75,-0.8);
\draw[fill=white] (4,0) circle (0.25);
\draw[fill=white](4.75,0.8) circle (0.25);
\draw (0.25,0) -- (0.75,0);
\draw[fill=white] (4.75,-0.8) circle (0.25);
\node at (1,-0.5) {$2$};
\node at (4,-0.5) {$p-1$};
\node at (5.5,0.8) {$p$};
\node at (5.5,-0.8) {$(1-)$};
\end{scope}
\begin{scope}[shift={(2,-2)}]
\fill[red] (0,0) circle (0.25);
\draw (0.25,0) -- (0.75,0);
\draw (1,0) circle (0.25);
\draw[dashed] (1.25,0) -- (2.75,0);
\draw (3,0) circle (0.25);
\draw (3.25,0) -- (3.75,0);
\draw (4,0) -- (4.75,0.8);
\draw (4,0) -- (4.75,-0.8);
\draw[fill=white] (4,0) circle (0.25);
\draw[fill=white](4.75,0.8) circle (0.25);
\draw (0.25,0) -- (0.75,0);
\draw[fill=white] (4.75,-0.8) circle (0.25);
\node at (1,-0.5) {$1$};
\node at (4,-0.5) {$p-2$};
\node at (5.5,0.8) {$p-1$};
\node at (5.5,-0.8) {$(1-)$};
\end{scope}
\end{tikzpicture} 
\end{center}
\caption{
Illustration of the mapping between the Bethe ansatz solutions of the light-cone six-vertex model (top) and those of the sine-Gordon model (bottom).
On the left panel, we represent typical configurations of rescaled Bethe roots. In the six-vertex model these assemble into strings (black) and real holes (red) on top of a Fermi sea of real roots (gray). In sine-Gordon the holes become the physical particles, and the strings become the magnonic strings, with a string order diminished by one. 
On the right panel we illustrate the mapping between the corresponding TBA Dynkin diagram, describing respectively the set of equations (\ref{eq:Dynkin2}) and (\ref{eq:Dynkin}).
}
\label{fig:strings}
\end{figure}
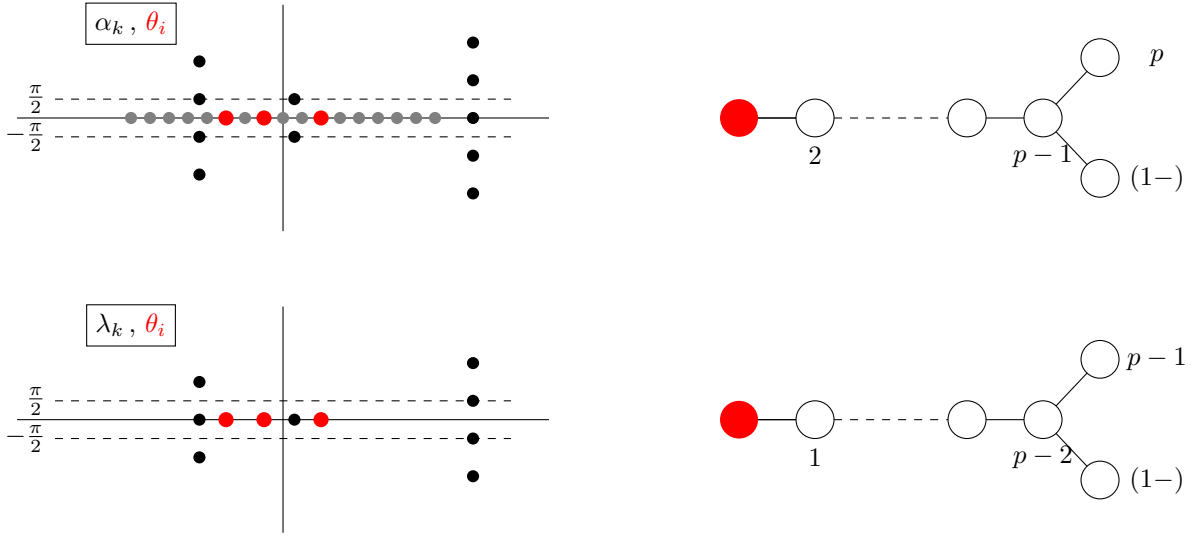

Going back to the above analysis for the repulsive case, we can notice at this stage that if the complex Bethe roots $\{ \alpha_k \}$ assemble into some regular patterns such as strings, so do the associated parameters $\{ \lambda_k \}$. This is illustrated in the left panel of figure \ref{fig:strings}, where we display the mapping between the set of holes of real roots and the complex roots of the light-cone six-vertex model, respectively onto the physical particles and magnonic rapidities of the sine-Gordon model. 
From this fact, is should be clear that in the case of a rational coupling $\frac{\beta^2}{8 \pi} = \frac{p}{p+1}$, the string Bethe equations of sine-Gordon (\ref{eq:Dynkin}) may be also directly derived from the string Bethe equations of the light-cone XXZ model. This is what we want to present now.

\subsection{String Bethe ansatz equations at rational values of the coupling} 
\label{sec:lightconestrings}

We now specify to rational values of the coupling  (\ref{eq:parameterp}), and, to keep the presentation simple, restrict further to the case where the parameter $p$ is an integer. In this case it is known (see Appendix \ref{sec:XXZ}) that the solutions of the Bethe equations (\ref{eq:BAEstag}) assemble into $m$-strings for $m=1,\ldots p$, as well as $(1-)$-strings, and we will show in this section that the sine-Gordon string Bethe equations can be directly derived in this case. 
 
In presence of the staggering, the Bethe equations for the densities of string centers and holes thereof, which we denote by $\{\rho^{\tiny \rm XXZ}_m(\alpha),\rho_m^{{\tiny \rm XXZ}, h}(\alpha)\}$ to avoid confusion with those introduced in section \ref{sec:stringsSG},  read (see Appendix \ref{sec:XXZ})
\be
\rho^{\tiny \rm XXZ}_m(\alpha) + \rho_m^{{\tiny \rm XXZ}, h}(\alpha)
  =  {a_m(\alpha+ \Theta/2)+a_m(\alpha- \Theta/2) \over 2}   - \sum_{n} \left(a_{m,n} \star \rho^{\tiny \rm XXZ}_n\right)(\alpha) ~~~ (m= 1,\ldots, p,(1-)) \,,
\label{eq:BetheTakahashiStag} 
\ee
or equivalently in Fourier space
 \be
\widehat{\rho}^{\tiny \rm XXZ}_m + \widehat{\rho}_m^{{\tiny \rm XXZ},h} =  \widehat{a}_m  \cos\frac{\omega \Lambda}{2} - \sum_{n} \widehat{a}^{\tiny \rm XXZ}_{m,n} \cdot \widehat{\rho}_n  \,,
\label{eq:BetheTakahashiFourierStag} 
\ee
where the different kernels are again those of equations (\ref{eq:BAEkernels},\ref{eq:BAEFourierkernels}).
As mentioned previously the ground state, or physical vacuum, corresponds to $\rho_1^{{\tiny \rm XXZ},h}= \rho^{\tiny \rm XXZ}_{m (\neq 1)}=0$, while the various holes or the strings of the type $m \neq 1$ play the role of excitations. One may then switch to the {\it physical equations}, where these various excitations act as sources over the physical vacuum, namely one uses the equation for $m=1$ to replace in the other equations $\widehat{\rho}^{\tiny \rm XXZ}_1$ by its expression in terms of $\widehat{\rho}_1^{{\tiny \rm XXZ},h}$ and $\widehat{\rho}^{\tiny \rm XXZ}_{m (\neq 1)},\widehat{\rho}_{m (\neq 1)}^{{\tiny \rm XXZ},h}$, yielding 
 \be
\widehat{\rho}^{\tiny \rm XXZ}_m + \widehat{\rho}_m^{{\tiny \rm XXZ},h} =    \frac{\widehat{a}_{m,1}}{1+\widehat{a}_{1,1}} \widehat{\rho}_1^{{\tiny \rm XXZ},h} 
- 
\sum_{n \neq 1}
\left(
  \widehat{a}_m 
  - 
  \frac{  \widehat{a}_{1,m}\widehat{a}_{m,1}}{1+\widehat{a}_{1,1}}
\right)
\widehat{\rho}^{\tiny \rm XXZ}_n
~~~~~ (m= 2,\ldots, p,(1-))
\,. 
\nonumber
\ee
These equations remain finite in the scaling limit $N \to \infty$, and can after some manipulations analog to those preceeding equation (\ref{eq:Dynkin}) be recast in the universal form  
\be
\rho_m^{{\tiny \rm XXZ},h} + \rho^{\tiny \rm XXZ}_m = \delta_{m,2} s \star \rho_1^{{\tiny \rm XXZ},h} + I_{m,n} s \star \rho_n^{{\tiny \rm XXZ},h}
 \qquad \qquad (m= 1,\ldots, p,(1-))
 \,,
\label{eq:Dynkin2}
\ee
where the source term $s$ is the one given in equation (\ref{eq:sdef}), the matrix $I$ is defined from its non-zero entries as 
\bea
I_{m,n} &=&  \delta_{m,n+1} + \delta_{m,n-1}  \qquad\qquad \mbox{($m=2, \ldots p-2$, $n=2, \ldots p-1$)} \nonumber \\
I_{p-1,n} &=&  \delta_{p-2,n} + \delta_{p,n}+\delta_{(1-),n}   \nonumber \\
I_{p,n} &=&  I_{(1-),n} = \delta_{p-1,n} \,, 
\eea
while we point out that in the sine-Gordon language the density of 1-holes $\rho_1^{{\tiny \rm XXZ},h}$ is related to the density of physical particles $\rho(\theta)$ (see below). 
As described in section \ref{sec:stringsSG} for the sine-Gordon Bethe ansatz equations, equations (\ref{eq:Dynkin2}) can be represented by the Dynkin diagram which has $I$ for incidence matrix, and which we represent on the top-right panel of figure \ref{fig:strings}.  
Up to a global shift of the string index, it is exactly the same as the sine-Gordon TBA diagram encoding (\ref{eq:Dynkin}), represented on the bottom-right panel of the same figure.

In other terms, the correspondence between the Bethe roots on the light-cone lattice and the original sine-Gordon TBA is the following : 
\begin{itemize}
\item The holes of 1-strings in the light-cone lattice become the physical particles of the sine-Gordon model, namely the solitons and antisolitons, with physical rapidities $\{\theta_k\}$.
The correspondence between the densities is 
\be
\rho(\theta) = \frac{N}{L}  \rho_1^{{\tiny \rm XXZ},h}(\theta) \,,
  \label{rhotheta}
\ee
where the proportionality factor stems from the fact that in sine-Gordon the densities are normalized with respect to the physical length $L$, while on the light-cone lattice they are normalized with respect to $N$.

\item For $2 \leq m \leq p-1$, the $m$-strings in the light-cone lattice become the $m-1$-strings of magnons in the sine-Gordon TBA, whose centers we parametrize by $\{\lambda^{(m-1)}_k \} = \{\alpha^{(m)}_k \}$.
The correspondence between the densities is 
\bea
\rho_{m-1}(\lambda) &=& \frac{N}{L}  \rho_{m}^{{\tiny \rm XXZ}}(\lambda)  \nonumber \\
\rho_{m-1}^h(\lambda) &=& \frac{N}{L}  \rho_{m}^{{\tiny \rm XXZ},h}(\lambda) 
  \,.
    \label{rhom}
\eea

\item The $(1-)$ strings in one case correspond to the $(1-)$ strings in the other, whose real parts we parametrize as $\{\lambda^{(1-)}_k \} = \{\alpha^{(1-)}_k \}$.
The correspondence between the densities is 
\bea
\rho_{(1-)}(\lambda) &=& \frac{N}{L}  \rho_{(1-)}^{{\tiny \rm XXZ}}(\lambda)  \nonumber \\
\rho_{(1-)}^h(\lambda) &=& \frac{N}{L}  \rho_{(1-)}^{{\tiny \rm XXZ},h}(\lambda) 
  \,.
  \label{rho1-}
\eea

\end{itemize} 

Let us mention in passing that for such rational values of the coupling the XXZ light-cone model can be reformulated as a restricted-solid-on-solid model, which in the technical language amounts to performing a {\it quantum group truncation} (see for instance \cite{DDV-RSOS}). For instance, in the present case of $p$ integer, this procedure would correspond to taking $\rho^{\tiny \rm XXZ}_{p-1} = \rho^{{\tiny \rm XXZ},h}_{p} = \rho_{(1-)}^{\tiny \rm XXZ} = \rho_{(1-)}^{{\tiny \rm XXZ},h} = 0$ in  (\ref{eq:Dynkin2}), or respectively $\rho_{p-1} = \rho^{h}_{p-1} = \rho_{(1-)} = \rho_{(1-)}^h = 0$ in  (\ref{eq:Dynkin}). 
The corresponding theories are the so-called restricted sine-Gordon models ${\rm RSG}(\beta^2/8\pi=p/p+1)$ \cite{RSG,DDV-RSOS}, and correspond to the $\Phi_{1,3}$ perturbations of the unitary minimal conformal models \cite{TBARSOS}.

Having elucidated the correspondence between the light-cone lattice Bethe ansatz solution and its sine-Gordon scaling limit, we now move on to the construction of the conserved charges which are the main object of this paper.

\section{Conserved charges from the light-cone discretization}
\label{sec:charges}

\subsection{Local lattice charges}

Before turning to the construction of new charges, we first review the well-known construction of local lattice charges, which in the scaling limit correspond to the integer Lorentz spin integrals of motion of the sine-Gordon theory \cite{SalResh}. 
In our conventions, these are defined from the logarithmic derivatives of the transfer matrix (\ref{eq:TXXZstaggered}) as
\be 
Q_{1,n}^{\pm} = 
 \left. \frac{\mathrm{d}^{n+1}}{\mathrm{d}\alpha^{n+1}}  \log
 \frac{T\left(\frac{\gamma}{\pi} i  \alpha \right)}{\sin \frac{\gamma}{\pi}(i\alpha +\frac{\pi}{2})^N} 
\right|_{\alpha=\Theta/2}
\pm
 \left. \frac{\mathrm{d}^{n+1}}{\mathrm{d}\alpha^{n+1}}  \log
 \frac{T\left(\frac{\gamma}{\pi} i  \alpha \right)}{\sin \frac{\gamma}{\pi}(i\alpha +\frac{\pi}{2})^N} 
\right|_{\alpha=-\Theta/2}
\,.
\label{eq:localcharges}
\ee

In particular, these include the momentum and Hamiltonian (\ref{eq:Pdef},\ref{eq:Hdef}), namely $P= \delta^{-1} Q_0^-$ and $H =\delta^{-1} Q_0^+$ up to some immaterial additional identity terms.

Since the matrices $T(u)$ form a family of mutually commuting operators, it is readily seen that the operators (\ref{eq:localcharges}) themselves commute with one another, and in particular with the Hamiltonian $H$. 
It can further be verified that these lattice operators are local, in the sense that they can be written as sums over lattice sites of densities with a finite support, namely 
\be
Q_{1,n}^{\pm}  = \sum_{i=1}^{N} q_{1,n}^{\pm[i,\ldots i+n]} \,,
\ee
where $q_{1,n}^{\pm [i,\ldots i+n]}$ acts non trivially only on the sites $i,\ldots i+n$ and as identity on the rest of the chain.

At this point, two remarks are in order : 
\begin{itemize}
\item  
The indices $\pm$ indicate the odd or even parity of the charges with respect to the left-right mirror symmetry of the model. 
This apparent doubling of the conserved charges is a feature of the light-cone construction, but does not mean that the light-cone discretization of a field should lead to more charges than the more traditional space-time discretization. In the latter time evolution is generally implemented by the row-to-row transfer matrix of a lattice model, whose logarithmic derivatives around the point of zero  spectral parameter lead to the local conserved charges. It can be easily seen that the parity-even charges (for instance the Hamiltonian) and the parity-odd charges (for instance the momentum) correspond to the even/odd moments, respectively (for an illustration, see \cite{milosz}).
 
\item Let us stress once again we have here taken non-conventional definitions the momentum and Hamiltonian, in order to include them in  our set of local charges. In the original light cone construction, neither the momentum (\ref{eq:Plc}) nore the Hamiltonian (\ref{eq:Hlc}) are local objects: while for the Hamiltonian this fact has to do with the light-cone geometry, the non-locality of the momentum operator is a completely generic feature. All of the considered objects are nevertheless part of the same commuting family, and this particular choice of momentum/Hamiltonian does not affect any of the conclusions presented in the following. 

\end{itemize}

\subsection{Quasilocal lattice charges}

As was first unveiled recently in the context of the XXZ chain in a series of papers by Prosen and collaborators (see in particular \cite{prosen-11,prosen-13,prosen-14,imp-15,Ilievski-GGE,zmp-16} as well as \cite{impz-16} for a review, and \cite{ppsa-14} by different authors and \cite{pv-16,pvc-16} for generalizations to chains of higher spin), the algebraic structures underlying quantum integrable lattice models may allow for more conserved charges to be built, which have a weaker (but still physically essential) form of locality called quasilocality. This term indicates that such charges can be written as sums of densities with arbitrarily large support, but with Hilbert-Schmidt norm decreasing exponentially with the length of the support. In the XXZ model, the charges relevant for the GGE construction were identified as the so-called unitary charges, built from the set of higher auxilliary spin transfer matrices \cite{Ilievski-GGE,zmp-16}. 
In this section we will follow quite closely the known construction for the XXZ case in order to define new, quasilocal conserved charges on the light-cone lattice introduced previously, the scaling limit of which will be taken in section \ref{sec:scalinglimitcharges}. 

As it is well known from the quantum integrability litterature, the staggered transfer matrix (\ref{eq:TXXZstaggered}) is actually the first member of a larger family of mutually commuting transfer matrices $\{T_j(u)\}_{j=1,2,\ldots}$. 
These are built by taking as the auxilliary space higher spin representations of the underlying $U_q(sl_2)$ quantum algebra ($q = \mathrm{e}^{i \gamma}$), namely
\be
T_j(u) = \mathrm{tr} \left(  L^{(j)}_N(u + i \Lambda/2) L^{(j)}_{N-1}(u - i \Lambda/2) \ldots L^{(j)}_1 (u-i \Lambda/2) \right)  \,,
\label{eq:Tj}
\ee
where the trace is over a spin-$j/2$ representation of $U_q(sl_2)$, namely some $(j+1)$-dimensional auxilliary space spanned by the vectors $\{|m\rangle\}_{m=-j/2,\ldots,j/2}$, and where the Lax operators $L_i^{(j)}$ are written as matrices acting respectively on each site of the chain, and whose entries are operators acting on the auxilliary space. Explicitely,
\be
L^{(j)} (u) = 
\left(
\begin{array}{cc}
\sin(u + \gamma (1/2+ S^{(j) z})) &  \sin(\gamma) S^{(j) -} \\
\sin(\gamma) S^{(j) +} &  \sin(u + \gamma (1/2-S^{(j) z})) \\
\end{array}
\right)
\label{eq:LaxXXZ}
\ee 
where the action of the generators $S^{(j) \alpha}$ is defined on the basis vectors of the auxilliary space as
\bea 
S^{(j) z} |m\rangle &=& m |m \rangle\,, \\
S^{(j) +} |m\rangle &=& \sqrt{[j/2+1+m]_q[j/2-m]_q} |m+1 \rangle\,, \\
S^{(j) -} |m\rangle &=& \sqrt{[j/2+1-m]_q[j/2+m]_q} |m-1 \rangle \,.
\eea 
In the set of equations above we have used the common notation for $q$-deformed numbers, $[x]_q = \frac{q^x - q^{-x}}{q-q^{-1}} = \sin(\gamma x) / \sin \gamma$.  
In particular it is easy to check that for $j=1$ the action of the matrix $L_i^{(1)}(u)$ on a site $i$ is equivalent to the matrix $R_{0,i}(u)$ of equation (\ref{Rmatrix}), and that $T_1(u)$ is simply the original transfer matrix $T(u)$ of equation (\ref{eq:TXXZstaggered}).

The transfer matrices (\ref{eq:Tj}) can further be shown to commute with one another, therefore they can be used to generate further charges commuting with the Hamiltionian. In particular, following \cite{Ilievski-GGE, Ilievski-stringcharge} we define the operators
\be
X_j (\alpha) =  \frac{\mathrm{d}}{\mathrm{d}\alpha}
\log
\left(
\frac{T_j\left(\frac{\gamma}{\pi} i  \alpha \right)}{\sin \frac{\gamma}{\pi}(i\alpha +j\frac{\pi}{2})^N} 
\right) \,.
\label{eq:Xjstagdef}
\ee
From these, we also define the following discrete sets of charges, 
\be
 Q_{j,n}^{\pm}  \equiv
  \left. \frac{\mathrm{d}^n}{\mathrm{d}\alpha^n}  \left( X_j\left(\alpha + \frac{\Theta}{2}\right) 
\mp 
X_j\left(-\alpha -  \frac{\Theta}{2}\right) 
  \right) \right|_{\alpha=0} 
  \,,
 \label{eq:Qjndef}
\ee 
which for $j=1$ coincide with the previously introduced local charges (\ref{eq:localcharges}). Here again the remark made above holds, on the meaning of the $\pm$ indices as the charges' parity with respect to the left-right spatial mirror symmetry of the model.

In Appendix \ref{app:quasilocal}, we prove that the operators (\ref{eq:Xjstagdef}), (and therefore also the charges (\ref{eq:Qjndef})) can be rewritten as a sum of densities with increasing support, namely 
\bea
X_j (\alpha) &=&  \sum_{r=1}^{N} \sum_{k=0}^{N/2} \mathcal{P}_{2 k}\left(  x^{ [r]}_j(\alpha) \right)   \,, \nonumber \\ 
Q_{j,n}^\pm &=&   \sum_{r=1}^{N} \sum_{k=0}^{N/2} \mathcal{P}_{2 k}\left(   q^{\pm [r]}_{j,n} \right)   \,,
\nonumber
\eea
where $\mathcal{P}_{2k}$ denotes a translation by $2k$ lattice sites, while the densities $x^{[r]}_j(\alpha)$ and $q^{\pm [r]}_{j,n}$ act non trivially only on $r$ consecutive sites, and as identity on the rest of the chain.

\begin{figure}
\begin{center}
\includegraphics[scale=0.55]{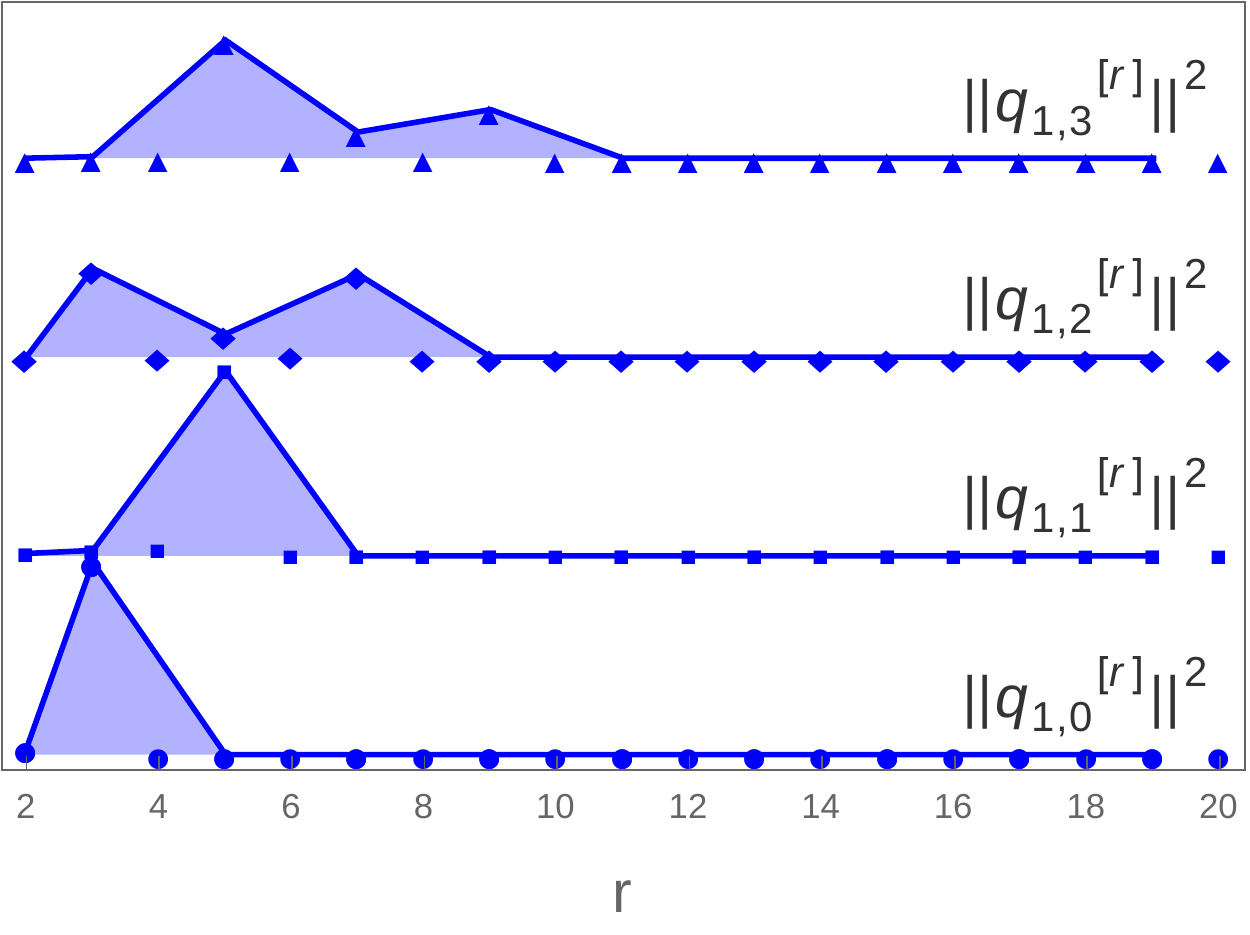}
\hspace{1cm}
\includegraphics[scale=0.55]{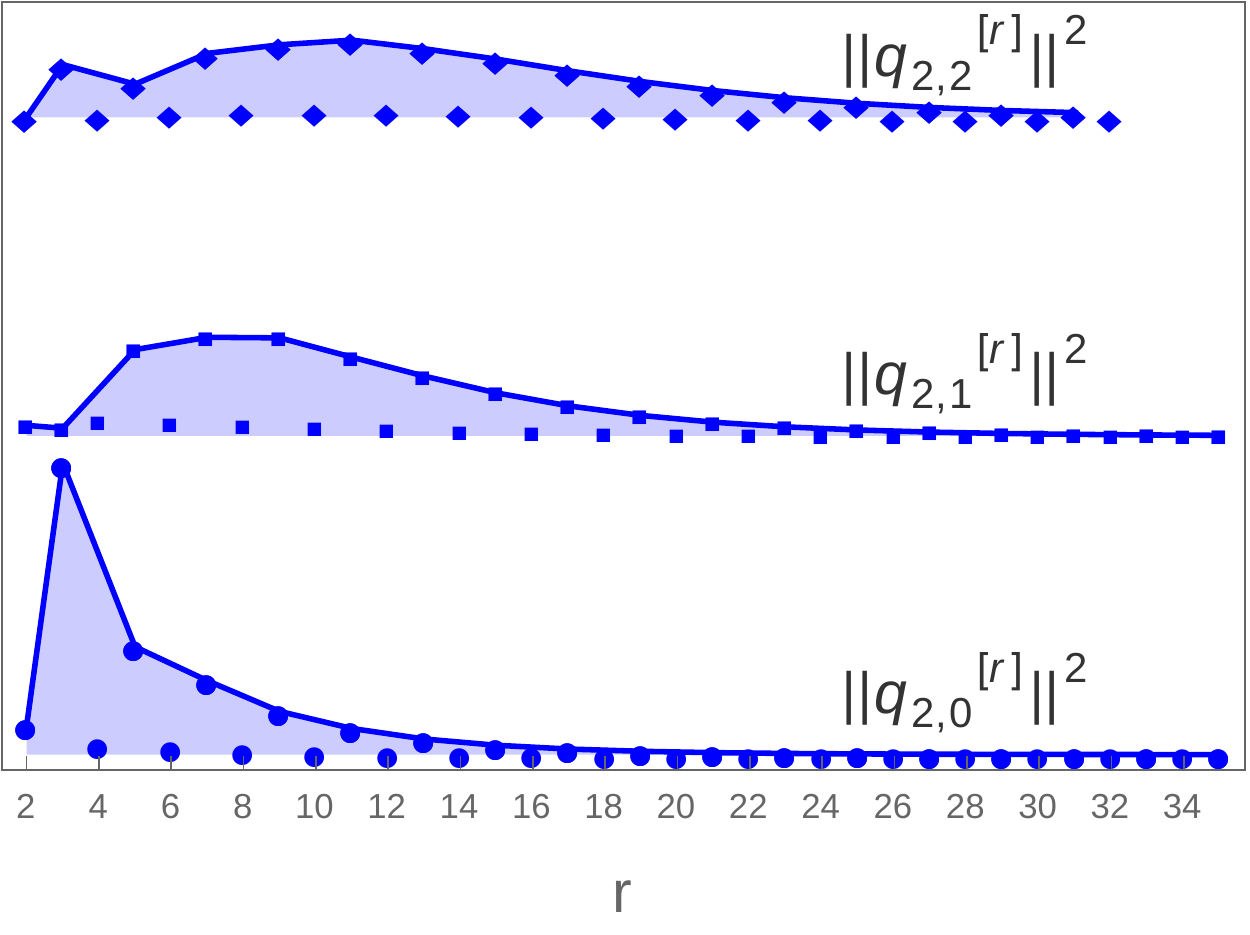}
\end{center}
\caption{
Squared Hilbert-Schmidt norms $||q_{j,n}^{\pm [r]}||^2_{\rm HS}$ of the finite support densities $q_{j,n}^{\pm [r]}$, as a function of $r$. The left and right panels correspond to $j=1$, and $j=2$ respectively. The data correspond to $\gamma=0.8$.
}
\label{fig:qnorms}
\end{figure}

For $\alpha$ lying in the so-called {\it physical domain}
\be
\mathcal{D}_{\gamma} = \left\{\alpha \in \mathbb{C} \big{|}~ |\mathrm{Im}(\alpha)| < \frac{\pi}{2} \right\}   \,,
\label{eq:physicaldomain}
\ee
it is further shown that the squared Hilbert-Schmidt norms of the densities $x_j^{[r]}(\alpha)$, as well as those of the densities $q_{j,n}^{\pm [r]}$, decrease with $r$ as 
\bea
||x_j^{[r]}(\alpha) ||^2_{\rm HS}  &\sim & \mathrm{e}^{-r / \xi_j(\alpha-\Theta/2)} 
\nonumber \\
||q_{j,n}^{\pm [r]}||^2_{\rm HS}  &\sim & \mathrm{e}^{-r / \xi_j(0)} \,,
\nonumber
\eea
where the characteristic lengths $\xi_j(\alpha)$ are given in equation (\ref{eq:xij}).
 We display on figure \ref{fig:correlationlengths} their values for $\alpha = 0$ and $\alpha=0.5$, and draw from there some conclusions, which essentially hold for any $\Lambda$ :    
\begin{figure}
\begin{center}
\includegraphics[scale=0.7]{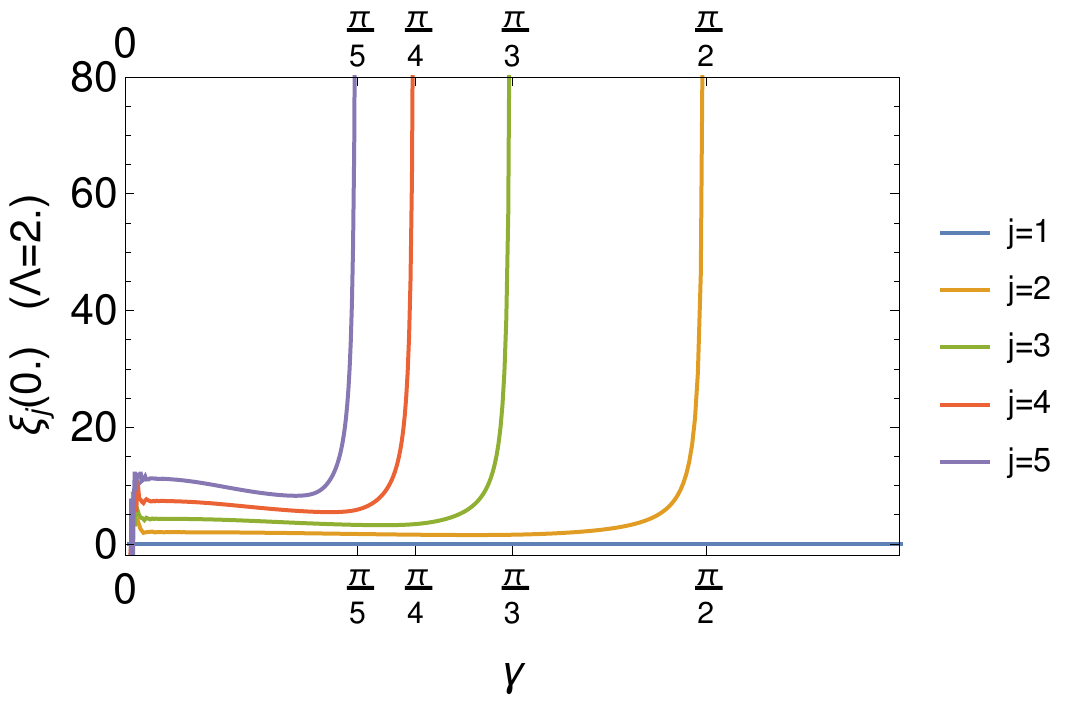}
\includegraphics[scale=0.7]{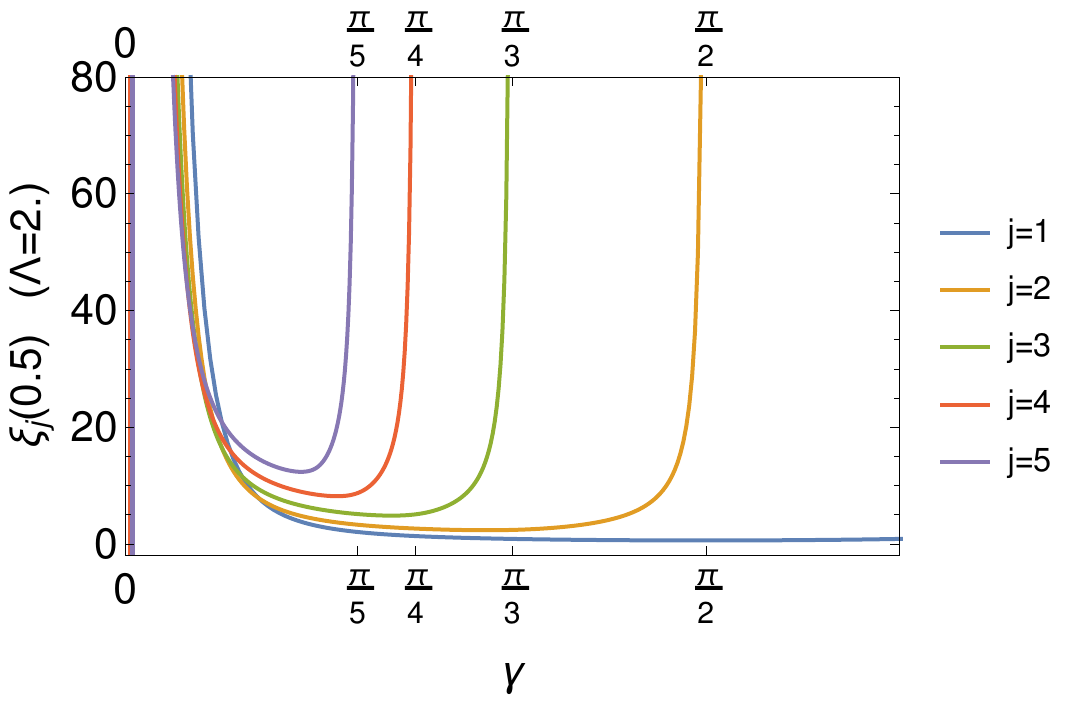}
\end{center}
\caption{
Characteristic lengths $\xi_j(\alpha)$ (in unit of lattice sites) of the exponentially decaying support of the quasilocal charges $X_j^{\pm}(\alpha)$, plotted as a function of $\gamma$ for $\Lambda=2$. 
}
\label{fig:correlationlengths}
\end{figure}
\begin{itemize}
\item $\xi_1(0)$ is zero, which goes along with the fact that the charges $Q_{1,n}^\pm$ of equation (\ref{eq:Qjndef}) are local (and not simply quasilocal). This can also be seen from the left panel of figure \ref{fig:qnorms}, where we represent directly the norms $||q_{1,n}^{\pm [r]}||^2$  for different values of $n$.
\item For $j\geq 2$, $\xi_j(\alpha)$ (for $\alpha$ in the physical domain) is finite for $\gamma < \frac{\pi}{j}$, and diverges at $\gamma = \frac{\pi}{j}$, which indicates that quasilocality breaks after this point. 
In other terms, the number of quasilocal charges is depends on the value of the coupling $\gamma$, a fact which has already been noticed by the authors of \cite{Ilievski-stringcharge} in the case of the homogeneous XXZ chain. Taking for instance $\gamma = \frac{\pi}{p+1}$ with $p$ integer, the set of linearly independent families of quasilocal charges corresponds to $\{X_j\}_{j=1,\ldots,p}$, which is precisely the set derived in \cite{Ilievski-stringcharge}.    
Let us mention in passing that for $p \leq 1$, which corresponds to the attractive regime of sine-Gordon, all the charges with $j\geq 2$ have a divergent decay length $\xi_j$. While we believe that quasilocal charges should also exist in this regime (as they do in the homogeneous case \cite{Ilievski-stringcharge}), the appropriate charge content requires a slightly different construction which we will not pursue here.

On the right panel of \ref{fig:qnorms}, we illustrate the property of quasilocality by representing the exponential decay of the norms $||q_{2,n}^{\pm [r]}||^2$ as a function of $n$ for different values of $n$. 
\end{itemize}

In conclusion, the quasilocal operators $X_j(\alpha)$ of \cite{Ilievski-GGE,Ilievski-stringcharge}, namely for $\alpha$ in the physical domain and $j$ suitably restricted to the set allowed by the value of the parameter $\gamma$, extend naturally to the imaginary staggered case ($\Lambda$ real). 
If these conditions are verified, we further check numerically that the corresponding transfer matrices $T_j$ satisfy the following {\it inversion relation} \cite{fe2-13,Ilievski-stringcharge}, 
\be
 \frac{T_j(i u - \gamma)}{\left( \frac{\sin\left(i u -\gamma \frac{j+1}{2}\right)}{\sin \gamma}  \right)^N}
 \cdot 
  \frac{T_j(i u )}{\left( \frac{\sin\left(i u +\gamma \frac{j+1}{2}\right)}{\sin \gamma}  \right)^N}
  = 1 + Y_j(i u) \,,
  \label{eq:inversionrelation}
\ee
with $||Y_j(i\alpha)||_{\rm HS} \to 0$ as $N\to \infty$. From there one can proceed as in \cite{Ilievski-GGE,Ilievski-stringcharge}, yielding for the eigenvalues of the operators $X_j(\alpha)$ the following additive expression in terms of the Bethe roots (since all the operators we consider are mutually commuting  we will use from this point the same notations for the operators and their eigenvalues),
\be
X_j(\alpha) = \sum_{\alpha_k} \phi_{j \over 2} (\alpha- \alpha_k ) \,,
\label{eq:XjBetheroots}
\ee
where the functions $\phi_{k}(\alpha)$ are defined as
\be
\phi_k(\alpha) =\frac{\mathrm{d}}{\mathrm{d}\alpha} \frac{i}{2}\log\frac{\sinh \left(\frac{1}{{p}+1} (\alpha - i k \pi) \right)}{\sinh \left(\frac{1}{{p}+1} (\alpha + i k \pi) \right)} \,.
\label{eq:phidef1}
\ee

In the next section, we will use this formula to investigate the scaling limit of the presently built local and quasilocal lattice charges.

\subsection{Scaling limit}
\label{sec:scalinglimitcharges}

The scaling limit, as described in section \ref{sec:lightconescaling}, is obtained by sending the lattice spacing $\delta$ to 0 and the number of lattice sites $N$ to infinity while keeping $L = N \delta$ fixed, while sending $\Lambda$ to infinity so as to keep fixed the value of the physical mass (\ref{eq:physicalmass}).
One easily sees from expression (\ref{eq:xij}) and the fact that the functions $f_j$ therein  $\to 1$ as $\Lambda \to \infty$ that the properties mentioned above for the characteristic lengths $\xi_j(\alpha)$ remain the same in the limit $\Lambda \to \infty$. As a result, in the domains where the former are finite, the associated physical lengths 
\be
\xi^{\rm phys}_j(\alpha)  = \delta ~ \xi_j(\alpha)
\nonumber
\ee
vanish, which means that the corresponding operators, in the scaling limit, act {\bf locally}. In the field theory context, we will therefore feel free to call these charges local, keeping in mind their quasilocal lattice counterparts.  
Our goal is now to investigate the action of these charges on the sine-Gordon particles, namely the solitons and antisolitons, and magnonic configurations. 
Anticipating on the results to be presented in this section, we want to derive expressions for the eigenvalues of the charges $Q_{j,n}^{\pm}$ on  a state specified by a set of physical rapidities $\{\theta_k \}$ and magnonic string rapidities $\{\lambda_k^{(m)}\}$ of the form (since all the operators we consider are mutually commuting, we will use the same notations for operators and their eigenvalues on Bethe states)
\be
Q_{j,n}^{\pm}  =  Q_{j,n}^{\pm {\rm vac}}  +    \sum_{\theta_k} q_{j,n}^{\pm}(\theta_k)
 +    \sum_{m}\sum_{\lambda_k^{(m)}} q_{j,n,m}^{\pm}(\lambda_k^{(m)})  
+ o(L) 
  \,,
 \label{eq:qjnrootsgeneric}
\ee
where $Q_{j,n}^{\pm {\rm vac}} $ is the extensive ($\propto N$) contribution of the light-cone lattice ground state.  
The residual term $o(L)$ in equation (\ref{eq:qjnrootsgeneric}) is expected to occur from taking the simultaneous limits $N,\Lambda \to \infty$. Its presence can actually be circumvented by taking the thermodynamic limit $L\to \infty$, where the particles and strings can be described by continuous densities $\rho(\theta), \rho_m(\lambda)$, and where the corresponding charges densities become
\be
{Q_{j,n}^{\pm}  -  Q_{j,n}^{\pm {\rm vac}}  \over L}  ~~ \stackrel{L\to \infty}{\longrightarrow}  ~~  \int \mathrm{d}\theta \rho(\theta) q_{j,n}^{\pm}(\theta) 
 +    \sum_{m}\int \mathrm{d}\lambda  \rho_m(\lambda) q_{j,n,m}^{\pm}(\lambda)   \,.
 \label{eq:qjnrootsdensities}
\ee
 In this limit the order of the limits $N \to \infty$ and $\Lambda \to \infty$, becomes irrelevant \footnote{We thank G\'{a}bor Tak\'{a}cs for pointing out this fact.}. We will therefore first send $N \to \infty$ while keeping $\Lambda$ finite, which will allow to derive a expression of the form (\ref{eq:qjnrootsgeneric}), {\it without the residual terms}.

In order to derive expressions such as (\ref{eq:qjnrootsgeneric}), our starting point is the expression  (\ref{eq:XjBetheroots}) of the operators $X_j(\alpha)$ in terms of the light-cone lattice Bethe roots $\{ \alpha_k \}$.   Using the string hypothesis as described previously, the sum (\ref{eq:XjBetheroots}) over all Bethe roots $\alpha_k$ can be recast as sum over the centers of strings of the different allowed types, namely, taking once again $\gamma = \frac{\pi}{p+1}$ with $p$ integer,
\be
X_j(\alpha) = \sum_{m} \sum_{\alpha^{(m)}_k} \phi_{j,m} (\alpha- \alpha^{(m)}_k )   \,,
\nonumber
\ee
where the functions $\phi_{j,m}$ are given by 
\bea
\phi_{j,m}(\alpha) &=& \sum_{k=1}^{\min(m,j)} \phi_{\frac{|m-j|-1}{2}+k}(\alpha)   ~~~~~~~~~ (m \leq p-1)  \\
\phi_{j,(1-)}(\alpha) &=& \phi_{j \over 2}(\alpha + i  (p+1)\pi/2)  \,.
\nonumber
\eea
In the $N \to \infty$ limit we switch to the densities $\rho^{\rm \tiny XXZ}_m$ and $\rho_m^{{\tiny \rm XXZ},h}$ for each type of string as described in section \ref{sec:lightconestrings}, yielding for $X_j(\alpha)$ an integral expression which we directly recast as an integral in Fourier space 
\be
X_j(\alpha) = \int_{-\infty}^{\infty} \mathrm{d} \omega~ \mathrm{e}^{i\omega \frac{\pi}{\gamma}\alpha}
\sum_{m}  \widehat{\phi}_{j,m}  \cdot \widehat{\rho}^{\rm \tiny XXZ}_m  \,,
\nonumber
\ee
Using the Bethe equations (\ref{eq:BetheTakahashiFourierStag}), one can make the replacement 
\be
\widehat{\rho}^{\rm \tiny XXZ}_1  =  \frac{\widehat{a}_{1} \cos \frac{\omega \Lambda}{2} }{1+\widehat{a}_{1,1}} - \frac{1}{1+\widehat{a}_{1,1}}
\widehat{\rho}_1^{{\tiny \rm XXZ},h}
-
\sum_{m \neq 1}
\frac{\widehat{a}_{1,m}}{1+\widehat{a}_{1,1}} \widehat{\rho}^{\rm \tiny XXZ}_m \,.
\nonumber
\ee
The first term in the right-hand side corresponds to the ground state density, or following section \ref{sec:lightconestrings} the sine-Gordon vacuum, while the further terms count the contributions from the various hole-like or complex-like excitations. 
From there we arrive at 
\be
X_j(\alpha)  =  X_j(\alpha)^{\rm vac} 
+ 
\int_{-\infty}^{\infty} \mathrm{d} \omega~ \mathrm{e}^{i\omega \frac{\pi}{\gamma}\alpha}
\left(  
\frac{-\widehat{\phi}_{j,1}}{1+\widehat{a}_{1,1}}  \widehat{\rho}_{1}^{{\tiny \rm XXZ},h}
+
\sum_{m\neq 1} \left( \widehat{\phi}_{j,m} - \frac{\widehat{a}_{1,m}\widehat{\phi}_{j,1}}{1+\widehat{a}_{1,1}} \right)\widehat{\rho}^{\rm \tiny XXZ}_m 
\right)
 \,,
 \label{eq:Xjvacplusexcitations}
\ee
where 
\be
  X_j(\alpha)^{\rm vac}  
  = 
  \int_{-\infty}^{\infty} \mathrm{d} \omega~ \mathrm{e}^{i\omega \frac{\pi}{\gamma}\alpha} 
\frac{\widehat{a}_{1} \cos \frac{\omega \Lambda}{2} }{1+\widehat{a}_{1,1}}  \widehat{\phi}_{j,1}
=
  \int_{-\infty}^{\infty} \mathrm{d} \omega~ \mathrm{e}^{i\omega \frac{\pi}{\gamma}\alpha} 
\frac{\sinh \omega\left( \frac{\pi}{2} - j \frac{\gamma}{2} \right) \cos \frac{\omega \Lambda}{2} }{2 \cosh \frac{\omega \gamma}{2}  \sinh \frac{\omega \pi}{2} } \,.
\ee

\subsubsection{Ultra local case ($j=1$)}

We start with the case $j=1$, namely with the ultralocal lattice charges $Q_1^{\pm}$. Note that these were already studied in \cite{SalResh}. 
Noticeably, for these charges the contributions of all kinds of strings vanishes from (\ref{eq:Xjvacplusexcitations}), so only holes contribute on top of the vacuum.  
On a configuration with holes of rapidities $\{ \theta_k \}$, we find
\be
X_1(\alpha)  =  X_1(\alpha)^{\rm vac} 
+ 
\sum_{ \theta_k} x_{1,{\rm holes}}(\alpha, \theta_k) \,,
\nonumber
\ee
where 
\be
x_{1,\mathrm{holes}}(\alpha,\theta) = \int_{- \infty}^\infty 
\mathrm{d}\omega
\mathrm{e}^{i\omega  \frac{\gamma}{\pi}(\alpha+\theta)}
\frac{1}{2 \cosh \frac{\omega\gamma}{2} } 
 = 
 \frac{\pi}{\gamma} \frac{1}{\cosh \left( {\theta+\alpha} \right)}
 \nonumber
\ee

It is then straightforward, using the definition (\ref{eq:Qjndef}), to obtain a similar expression for the charges $Q_{1,n}^{\pm}$. In the scaling limit, these can be written in terms of the physical mass $M$ (\ref{eq:physicalmass}) and rapidities $\theta_k$ as 
\be
Q_{1,n}^{\pm} =  Q_{1,n}^{\pm \rm vac} 
+ 
\sum_{\theta_k} q^{\pm}_{1,n}(\theta_k) \,,
\nonumber
\ee
where the functions $q^{\pm}_{1,n}$ can be written as the following series expansions 
\be
q_{1,n}^{\pm}( \theta) =
\frac{4 \pi}{\gamma}
\sum_{k=0}^{\infty}
(-1)^k (2k+1)^n
(\delta M)^{2k+1}
{c}_{\pm} \left( (2k+1)\theta \right) \,. \label{spinsultralocal}
\ee
Here and for the following we have introduced the functions
\bea
c_{+}(\theta) &=&  \cosh \theta  \nonumber \\
c_{-}(\theta) &=&  \sinh \theta    \,.
\nonumber
\eea

In particular, keeping only the leading term in the scaling limit $\delta \to 0$, we recover for the energy and momentum the usual expression for relativistic particles of mass $M$ and rapidity $\theta$
\bea
e(\theta)  = \delta^{-1} q_{1,0}^{+}(\theta)  &\propto& M \cosh \theta \nonumber \\
p(\theta) =  \delta^{-1} q_{1,0}^{-}(\theta) &\propto& M \sinh \theta \,.
\nonumber
\eea
Noticeably, there is no contribution for the magnons, which agrees with the fact that these carry no energy or momentum. This observation extends to the charges of higher Lorentz spin which can be built out of the $q_{1,n}^{\pm}$ by taking appropriate linear combinations \cite{SalResh}. For instance, the operators $\delta^{-3} (Q_{1,0}^{\pm} - Q_{1,1}^{\pm})$ are easily seen to be the sum of contributions of the form $M^3 c_{\pm}(3 \theta)$ for each individual particle. More generally, one may construct conserved charges whose single particle contributions are of the form $M^{2k+1}\cosh (2k+1)\theta$, $M^{2k+1}\sinh (2k+1)\theta$, this for any odd-integer value $2k+1$. 

Here we would like to point out that one can easily read off the Lorentz spin of the conserved charges in the scaling limit, from the expression (\ref{spinsultralocal}). A conserved charge whose eigenvalue on a one-particle state is of the form $c_{\pm}(s\theta)$, transforms as an $s$-rank tensor under a Lorentz transformation.  The best known example is the $s=1$ case, which corresponds to the conserved energy and momentum operators, which transform as a vector. It is then easy to see that the set of charges $Q_{1,n}^\pm$ yield in the scaling limit, a set of charges with odd-integer Lorentz spins $s=2k+1$. We will see in the next section, using the same argument, that the quasilocal lattice charges yield in the scaling limit a set of charges with fractional Lorentz spin.

\subsubsection{Quasilocal case ($j\geq 2$)}

For the charges ${X}_j$ with $j \geq 2$, in contrast with the case $j=1$, a non zero contribution comes from the configurations of the XXZ strings. Indicating the centers of $m$-strings by $\{\lambda^{(m-1)}_k \} = \{\alpha^{(m)}_k \}$ (respectively, $\{\lambda^{(1-)}_k \} = \{\alpha^{(1-)}_k \}$ for $(1-)$-strings), we indeed find 
\be
X_j(\alpha)  =  X_j (\alpha)^{\rm vac} 
+ 
\sum_{\theta_{k}} x_{1,{\rm holes}} (\alpha,\theta_k) 
+ 
\sum_{m \neq 1}\sum_{\alpha^{(m)}_{k}} x_{1,m} (\alpha,\alpha^{(m)}_k)
\,,
\nonumber
\ee
where the contributions of all kinds are 
\bea
x_{j,\mathrm{holes}}^{\pm}(\theta,\alpha) &=&  \int_{- \infty}^\infty 
\mathrm{d}\omega  \mathrm{e}^{i \frac{\gamma}{\pi} \gamma (\alpha+\theta)} 
\frac{\sinh\omega \left( \frac{\pi}{2} - j\frac{\gamma}{2} \right)}{2 \cosh \frac{\omega\gamma}{2} \sinh\omega \left( \frac{\pi}{2} -\frac{\gamma}{2} \right) }
\nonumber \\
x_{j,m}^{\pm}(\alpha^{(m)},\alpha)  &=& \int_{- \infty}^\infty \mathrm{d}\omega 
\mathrm{e}^{i\omega \frac{\gamma}{\pi}(\alpha+\alpha^{(m)})}  
\frac{\sinh\omega\left( \frac{\pi}{2} - j \frac{\gamma}{2} \right)}{\sinh \frac{\omega \pi}{2}}
\left(
\frac{\sinh \frac{m\omega \gamma}{2}}{\sinh \frac{\omega \gamma}{2} }
-
\frac{\sinh\omega\left( \frac{\pi}{2} - m \frac{\gamma}{2} \right)}{\sinh\omega\left( \frac{\pi}{2} -  \frac{\gamma}{2} \right)}
\right) ~~~~ (m \leq j \leq p)
\nonumber \\
x_{j,m}^{\pm}(\alpha^{(m)},\alpha) &=& \int_{- \infty}^\infty \mathrm{d}\omega 
\mathrm{e}^{i\omega  \frac{\gamma}{\pi}(\alpha+\alpha^{(m)})} 
\frac{\sinh\omega\left( \frac{\pi}{2} - m \frac{\gamma}{2} \right)}{\sinh \frac{\omega \pi}{2}}
\left(
\frac{\sinh \frac{j \omega \gamma}{2}}{\sinh \frac{\omega \gamma}{2} }
-
\frac{\sinh\omega\left( \frac{\pi}{2} - j \frac{\gamma}{2} \right)}{\sinh\omega\left( \frac{\pi}{2} -  \frac{\gamma}{2} \right)}
\right)~~~~ (j \leq m \leq p)
\nonumber \\
x_{j,(1-)}^{\pm}(\alpha^{(1-)},\alpha) &=& \int_{- \infty}^\infty \mathrm{d}\omega 
\mathrm{e}^{i\omega \frac{\gamma}{\pi} (\alpha+\alpha^{(1-)})} 
\frac{1}{\sinh \frac{\omega \pi}{2}}
\left(
\frac{\sinh\omega \gamma\left(\frac{j-1}{2}\right)
}
{\sinh\omega\left( \frac{\pi}{2} -  \frac{\gamma}{2} \right)}
\right) \,.
\nonumber
\eea

All these integrals can be computing by residues (note that $\alpha$ belonging to the physical domain (\ref{eq:physicaldomain}) is a necessary and sufficient condition for all the integrals to converge), resulting for $\Lambda >0$ in a series expansion in negative exponentials of $\Lambda$.  
From there we can as in the $j=1$ case derive expressions of the charges $Q_{j,n}^{\pm}$ in the scaling limit in terms of the physical rapidities $\{\theta_k \}, \{\lambda^{(m-1)}_k \}$. Namely these have the form (\ref{eq:qjnrootsgeneric}), where the contributions of various kinds are found as 
\bea
q_{j,n}^{\pm}(\theta) 
&=&
\frac{4 \pi}{\gamma}
\sum_{k=0}^{\infty}
(-1)^k(2k+1)^n
\frac{\sin \left( \frac{\pi(\pi-j\gamma)}{2 \gamma} (2k+1) \right) }{\sin \left( \frac{\pi(\pi-\gamma)}{2 \gamma} (2k+1) \right) }
(\delta M)^{2k+1}
c_{\pm}  \left( (2k+1)\theta \right) 
 \nonumber \\
 &+& 
 \frac{4 \pi}{\pi-\gamma}
\sum_{k=0}^{\infty}
(-1)^k \left(\frac{2 k \gamma}{\pi - \gamma} \right)^n
\frac{\sin \left( \frac{\pi(\pi-j\gamma)}{\pi- \gamma} k  \right) }{\sin \left( \frac{\pi(\pi-\gamma)}{\pi- \gamma} k \right) }
(\delta M)^{\frac{2 k \gamma}{\pi - \gamma}}
c_{\pm}  \left(\frac{2 k \gamma}{\pi - \gamma}\theta \right)  
\nonumber \\
q_{j,n,m}^{\pm}(\lambda) 
&=&
 \frac{4 \pi}{\pi-\gamma}
\sum_{k=0}^{\infty}
(-1)^k\left(\frac{2 k \gamma}{\pi - \gamma} \right)^n
\frac{\sin \left( \frac{\pi(\pi-j\gamma)}{\pi- \gamma} k  \right) 
\sin \left( \frac{\pi(\pi-(m+1)\gamma)}{\pi- \gamma} k  \right) 
}{\sin \left( \frac{\pi^2 }{\pi- \gamma} k \right) }
(\delta M)^{\frac{2 k \gamma}{\pi - \gamma}}
c_{\pm}  \left(\frac{2 k \gamma}{\pi - \gamma}\lambda \right)  
\nonumber \\
& & 
\hspace{11cm} (m \leq p-1)
\nonumber \\
q_{j,n,(1-)}^{\pm}(\lambda) 
&=&
 \frac{4\pi}{\pi-\gamma}
\sum_{k=0}^{\infty}
(-1)^k\left(\frac{2 k \gamma}{\pi - \gamma} \right)^n
\sin \left( \frac{(j-1) \pi \gamma}{\pi- \gamma} k  \right) 
(\delta M)^{\frac{2 k \gamma}{\pi - \gamma}}
c_{\pm}  \left(\frac{2 k \gamma}{\pi - \gamma} \lambda\right)   \,.  
\nonumber \\
\label{spinsquasilocal}
 \eea
 
 Let us insist on the fact that while all of the above results were derived for integer values of the parameter $p$, there is in principle no obstruction to proceed similarly for arbitrary rational values of $p>1$, which cover densely the sine-Gordon repulsive regime. For a given value of $p$, the string content of the Bethe ansatz solutions and the set of quasilocal charges will be different, but it will remain possible to write for the latter expressions of the form (\ref{eq:qjnrootsgeneric}).

Now, there are two main important properties of the eigenvalues (\ref{spinsquasilocal}) that we wish to point out. The first is that contrary to the ultralocal charges, the quasilocal ones have nonzero eigenvalues for auxiliary particles (strings). The second important property is that, just as we did for the eigenvalues of the ultralocal charges, we can read off the Lorentz spins of the quasilocal charges in the continuum limit, from the expressions (\ref{spinsquasilocal}). Particularly, from the rapidity-dependent factors of $c_\pm\left(\frac{2k\gamma}{\pi-\gamma}\theta\right)$, we see that these charges have the Lorentz transformation properties of a set of charges with fractional, and coupling-constant dependent Lorentz spin. 
In terms of the parameter $p$ of equation(\ref{eq:parameterp}), the set of new Lorentz spins reads
\be 
s = \frac{2 \pi k}{p} \,, \qquad k =1,2,\ldots    \,.
\label{eq:Lorentzspins}
\ee

It is not the first time that conserved charges with a fractional Lorentz spin are found in the sine-Gordon model \cite{Bernard1,Bernard2,Russians}. In the next section, we will review the previously existing constructions and emphasize what are their common features and differences with ours. 
complex roots

\subsection{Comparison with the previously known charges}

\subsubsection{Non-local and non-commuting quantum group charges}

In \cite{Bernard1,Bernard2}, Bernard and LeClair have constructed a set of non-local charges associated with a set of non-local conserved currents in the sine-Gordon model. The non-locality of these currents  is manifest from the fact that their definition at a space-time point $(x,t)$ involves attaching a ``tail" running from infinity to $(x,t)$, analogous to that encountered in the definition of disorder operators e.g. in the Ising model, and which is responsible for non-trivial braiding relations between the currents and non-trivial commutation relations between the charges. Unlike the conserved charges we have found in the previous sections, the nonlocal charges of Bernard and Leclair do not commute with each other. In fact, these charges have been shown to be elements of the quantum  algebra $U_{\tilde{q}}(\widehat{sl(2)})$, where, in our notations, $\tilde{q} = e^{-\frac{2 i \gamma}{\pi-\gamma}}$.
 In appendix \ref{app:othercharges}, we show that the charges constructed in \cite{Bernard1,Bernard2} have Lorentz spin 
\be 
 s =  \frac{2 \gamma}{\pi - \gamma} =  \frac{2}{p} \,,
\label{eq:LorentzspinsBL}
\ee
which coincides precisely with the lowest fractional spin of the charges considered in this paper, equation (\ref{eq:Lorentzspins}). 
It is further suggested in  \cite{Bernard1,Bernard2} that the higher level charges may similarly be generated from operator product expansions. These will have for Lorentz spins integer multiples of (\ref{eq:LorentzspinsBL}), matching perfectly the spin content of our charges. 

Despite all this, the charges of \cite{Bernard1,Bernard2} are highly non-local by definition and non commuting, and are therefore very different than the ones built in this work. In fact, the lattice equivalent of such charges has been proposed in \cite{BernardFelder}, and may be related to the recently devised discretely holomorphic observables in integrable lattice models \cite{Ikhlef}.  
At this moment we do not fully understand whether the coincidence of values of spins points to some deeper mathematical connection between the two sets of charges. In particular, one likely scenario is that our conserved charges may form the maximal Abelian subalgebra of the quantum group charges. The locality of such charges, from the perspective of \cite{Bernard1,Bernard2}, would then result from the cancellation of contributions resulting from integrating the non-local field $\Theta$ (see Appendix \ref{app:othercharges}).

\subsubsection{Non-local, commuting charges from transfer matrices of conformal field theory}
\label{sec:BLZ}

Another interesting development regarding the existence of fractional spin conserved charges in sine-Gordon came from Bazhanov, Lukyanov and Zamolodchikov in Ref. \cite{Russians,Russians2}. Their approach consists on finding representations of the transfer matrices and $Q$-functions of conformal field theories (CFT's) directly in the continuum field formalism, without reference to an underlying lattice system. From this representation of the transfer matrices, a set of non-local commuting charges can be obtained, in a procedure similar to ours. The relation between these CFT charges, and conserved charges in sine-Gordon was further elucidated in the review \cite{negro}, and some aspects of it are reviewed in Appendix \ref{app:othercharges}. 

What we see in particular is that the construction of Bazhanov {\it et.al} leads to a set of conserved charges in sine-Gordon with fractional Lorentz spin that matches exactly that of the charges from our construction. By construction these charges also all commute with each other, as do the charges of the present paper. While we are not able at this stage to precisely relate our charges to those of Bazhanov, {\it et.al}, it seems likely that some close connexion should exist between the two sets,  for instance that the two be linearly related. One possible way to elucidate this relation could be to relate the field-theoretical transfer matrices of \cite{Russians} to the lattice transfer matrices of this work, using the fact that these satisfy the same set of functional equations\footnote{This was suggested to us by G\'{a}bor Tak\'{a}cs}. Both families of transfer matrices are indeed solutions of  hierarchy of fusion relations (T system), together with a TQ equation \cite{Russians2,negro}. Based on uniqueness properties, it would therefore be enough that the $Q$ operator of the field theory and that of the light-cone discretization coincide in order to conclude that the continuum limit of the lattice transfer matrices should coincide with those of the field theory. At this stage, we have however been unable to derive a tractable expression scaling limit of the lattice $Q$ operator which would make it comparable with that of \cite{Russians2,negro}. 
Another direction could be to look at the conformal limit of the charges, as was done in \cite{Russians} for the local ones (derived from the fundamental transfer matrix)\footnote{This was suggested to us by Hubert Saleur.}. Namely, the field-theory charges of \cite{Russians} are written as combinations of powers of the holomorphic and antiholomorphic components of the stress energy tensor, from which one can extract universal contributions to the scaling of these charges as a function of the size $L$ (see equation (40) in \cite{Russians}). 
These universal contributions may in turn be compared to expressions of the form (\ref{spinsquasilocal}), namely on a given sine-Gordon scattering state sums of the form $\sum \mathrm{e}^{s\theta}$ over the different particles may be evaluated through massless TBA \cite{fendley}, and linear combinations thereof may be taken appropriately so as to match the above-mentioned universal terms. We leave this aspect to future investigation.

A striking aspect is the manifest non-locality of the charges (\ref{chargesBLZ}), while ours become local in the scaling limit. Were it to be proven that a linear relation exists between the two, as advocated above, this would indicate that the former are, despite all appearances, local as well, a fact which it has yet been impossible to prove from the field-theoretical construction.

\section{Construction of the GGE}
\label{sec:GGESG}

In the previous section, we have built a family of quasilocal conserved operators $X_j(\alpha)$ on the light-cone lattice, which we have argued to become local in the sine-Gordon scaling limit for a properly chosen set of values of $j$ and for $\alpha$ in the physical domain (\ref{eq:physicaldomain}). In the homogeneous XXZ case, analogously built quasilocal charges were shown \cite{Ilievski-GGE, Ilievski-stringcharge,pvc-16} to determine completely the densities of all kinds of strings needed to describe the steady-state following a quantum quench. 

In this section, we will show how similar conclusions may be drawn in the sine-Gordon case, namely how the operators $X_j(\alpha)$ are {\it in principle} enough to determine the densities of physical particles and holes thereof, $\rho(\theta), \rho^{h}(\theta)$ as well as those associated with the various kinds of magnonic strings, $\{\rho_m(\lambda), \rho_m^{h}(\lambda)\}$.  We will then sketch how the construction of a complete GGE may be achieved in terms of the charges $Q_j^{\pm}$.

\subsection{From the operators $X_j(\alpha)$ to the string densities}

The results of this section follow mostly those of \cite{Ilievski-GGE, Ilievski-stringcharge} for the homogeneous XXZ case, which take the form of a set of linear relations between the densities $\{\rho_m,\rho_m^h\}$ of a given eigenstate and the associated expectation values of the operators $X_j(\alpha)$. Namely,
 \bea
\rho_j(\alpha) &=&  \frac{1}{N}X_j\left(\alpha+i\frac{\pi}{2}\right)+ \frac{1}{N}X_j\left(\alpha-i\frac{\pi}{2}\right)- \frac{1}{N}X_{j+1}\left(\alpha\right)- \frac{1}{N}X_{j-1}\left(\alpha\right)\,,
\nonumber
\\
\rho_{j}^{h}( \alpha) &=& a_j(\alpha)-  \frac{1}{N}X_j \left(\alpha+i\frac{\pi}{2}\right) -  \frac{1}{N}X_j \left(\alpha-i\frac{\pi}{2}\right) \,,
\label{eq:rhoX}
\eea
as $N\to \infty$ (note the different normalization with respect to \cite{Ilievski-stringcharge}).
The relations (\ref{eq:rhoX}), dubbed ``string-charge duality'' in \cite{Ilievski-stringcharge}, are readily derived from the inversion relation (\ref{eq:inversionrelation}) and hold under the following restrictions : 
\begin{itemize}
\item The operators $X_j(\alpha)$ are quasilocal within the physical strip (\ref{eq:physicaldomain}). This restricts the validity of the equations (\ref{eq:rhoX}) to a set of values of the parameter $j$ for which the former property is true, but which is further shown in \cite{Ilievski-stringcharge} to match perfectly the string content of the theory.  
 For instance, in the case where the XXZ coupling $\tilde{p}$ is some positive integer (in the notations of Appendix \ref{sec:XXZ}), the set of possible strings is $m=1,2, \ldots \tilde{p}, (1-)$, while the set of quasilocal charges corresponds to $j=1, 2, \ldots \tilde{p}$. 
However, as explained in \cite{Ilievski-stringcharge}, the strings of type $\tilde{p}$ and $(1-)$ do not carry independent dynamical information, so the set of equations (\ref{eq:rhoX}) is enough to fix all the information about all densities.  

\item The meromorphic functions $X_j(\alpha)$ must not have poles on the physical strip (\ref{eq:physicaldomain}). In practice this fact may be checked directly from the evaluation of the charges $X_j(\alpha)$ on the initial states of interest, and turns out to be verified for the various types of quenches considered in the XXZ litterature \cite{Ilievski-GGE, Ilievski-stringcharge,pvc-16}. 
\end{itemize}
 
In the present, staggered case, the inversion relation (\ref{eq:inversionrelation}) has been checked numerically from finite size data, and was found to hold, at a given value of the XXZ coupling, for the same set of values of $j$ as in the homogeneous case. 
From there one can follow the same steps as in the homogeneous case \cite{Ilievski-GGE, Ilievski-stringcharge}, resulting in the following relations between the light-cone charges and densities :
  \bea
\rho^{\tiny \rm XXZ}_j(\alpha) &=&  \frac{1}{N}X_j\left(\alpha+i\frac{\pi}{2}\right)+ \frac{1}{N}X_j\left(\alpha-i\frac{\pi}{2}\right)- \frac{1}{N}X_{j+1}\left(\alpha\right)- \frac{1}{N}X_{j-1}\left(\alpha\right)\,,
\nonumber
\\
\rho_{j}^{{\tiny \rm XXZ}, h}( \alpha) &=& \frac{a_j(\alpha+\Theta/2)+ a_j(\alpha-\Theta/2)}{2}-  \frac{1}{N}X_j \left(\alpha+i\frac{\pi}{2}\right) -  \frac{1}{N}X_j \left(\alpha-i\frac{\pi}{2}\right) \,,
\label{eq:rhoXstag}
\eea

In the scaling limit the source term in the second equation of (\ref{eq:rhoXstag}) vanishes. In terms of the densities  of sine-Gordon physical particles and magnons, obtained from (\ref{rhotheta},\ref{rhom},\ref{rho1-}), this becomes
  \bea
\rho(\theta) &=& -  \frac{1}{L}X_{1} \left(\theta+i\frac{\pi}{2}\right) -  \frac{1}{L}X_{1} \left(\theta-i\frac{\pi}{2}\right) \,,
\nonumber 
\\
\rho_{j}(\lambda) &=&  \frac{1}{L}X_{j+1}\left(\lambda+i\frac{\pi}{2}\right)+ \frac{1}{L}X_{j+1}\left(\lambda-i\frac{\pi}{2}\right)- \frac{1}{L}X_{j+2}\left(\lambda\right)- \frac{1}{L}X_{j}\left(\lambda\right)\,,
\nonumber
\\
\rho_{j}^{h}( \lambda) &=& -  \frac{1}{L}X_{j+1} \left(\lambda+i\frac{\pi}{2}\right) -  \frac{1}{L}X_j \left(\lambda-i\frac{\pi}{2}\right) \,.
\label{eq:rhoXSG}
\eea
The set of equations (\ref{eq:rhoXstag}) provides linear relations between the expectation values of the conserved operators $X_j(\alpha)$ and the densities of sine-Gordon physical particles and magnons. 
The densities describing the steady state resulting from a quantum quench from an initial state $|\Psi_0\rangle$ may therefore in principle be obtained after evaluating the expectation values of the operators $X_j$ on $|\Psi_0\rangle$. 
It is however worth to insist that while this procedure has been successfully applied to the XXZ case in (\cite{Ilievski-GGE, Ilievski-stringcharge,pvc-16}), its implementation in the field-theoretical setup requires some more work. Indeed, while the charges $X_j(\alpha)$ have a well-defined operator formulation on the lattice, we have in the sine-Gordon language only provided their action on the various Bethe ansatz eigenstates. This is clearly not enough to compute their action on some arbitrary initial state, for instance an alternation of solitons and antisolitons with increasing rapidities. 
In order to make progress in this direction, it would be useful to confirm the relationship proposed in section (\ref{sec:BLZ}) between the charges $X_j(\alpha)$ and the ``non-local'' charges of Bazhanov, Lukyanov and Zamolodchikov \cite{Russians}, which are defined directly in the field-theoretical setup and whose action on arbitrary states should be more easily understood. We hope to be able to report on this soon.

\subsection{The complete GGE}

Despite the practical difficulties mentioned in the previous section, our construction suggests that the quasilocal light-cone operators $X_j(\alpha)$ contain all the information necessary to fully characterize a steady state in the sine-Gordon model. That is, if one can measure the expectation values of $X_j(\alpha)$ for every value of $\alpha$ and $j$, one can recover all the particle density distributions. 

In the GGE description of steady states, one assumes that expectation values of local operators can be computed with an ensemble that includes only local or quasilocal conserved charges. In the following, we will describe the set of local and quasilocal charges that must be included in the GGE to accurately compute local observables in sine-Gordon. As we will see, the scaling limit of the lattice charges $Q_{j,n}^{\pm}$ requires some careful prescription, which follows the spirit of the construction presented in \cite{milosz}, where the complete set of conserved charges for some field theories with diagonal scattering was found. We will review this case first, and come back to sine-Gordon afterwards.

\subsubsection{Review of the construction for diagonal scattering}

 As we have discussed previously, the diagonal scattering case corresponds to theories where there are no magnons and strings, but only physical particles. 
The prime example of diagonal scattering considered in \cite{milosz} is the transverse field Ising chain on a lattice with spacing $\delta$, and total number of sites, $N$, whose continuum limit describes free Majorana fermions. There is a discrete set of ultralocal conserved charges with support on an integer number of lattice sites, that goes from $1$ to $N$. The new insight of \cite{milosz} is that there are two different ways to take the continuum limit of these charges, which are described pictorially on the two left panels of figure \ref{fig:limitingprocedure}.

The standard, previously known charges in the continuum limit are given by considering  conserved charges $\{Q_n\}_{n  \in \mathbb{N}}$ on the lattice with support on $n$ sites, then taking the continuum limit, $\delta\to 0,\,N\to\infty$, while letting $n$ be a finite number, such that $n \delta\to 0$. This procedure yields a set of local conserved charges with integer Lorentz spin, and with support on a vanishingly small region of space. The proposal of Ref. \cite{milosz} is that this limiting procedure is incomplete, and this set of local charges is not enough to completely determine the particle density distributions. Indeed, these local charges only capture information about the moments of the continuous density distributions, $\rho(\theta)$, and certain counter-examples were presented in \cite{milosz} where it was shown that knowledge of these moments is not necessarily sufficient to reproduce the full density distributions.

In order to solve this problem, a complementary limiting procedure was suggested which consists on taking charges on the lattice with support on $n$ sites, and letting $\delta\to0,\,N\to\infty$, while taking $n\to\infty$, keeping $\alpha \equiv n \delta$ finite and nonzero. This leads to a continuous set of ``semilocal"\footnote{We denote by ``semilocal'' what is actually dubbed ``quasilocal" in Ref. \cite{milosz}, in order to avoid confusion with the presently used, and generally accepted, notion of quasilocality. In particular the semilocal charges of \cite{milosz} have finite support in a region of space.} charges $\{Q(\alpha)\}_{\alpha \in \mathbb{R}_+}$ in the field theory, which have support on the region of space with size $\alpha=n \delta$. This continuous set of semilocal charges was shown in \cite{milosz} to be in one-to-one correspondence with the momentum occupation modes, $I(\theta)$, and therefore to completely fix the stationary state.
Resultingly, the GGE density matrix for such models was proposed in \cite{milosz} to be of the form
\be 
\varrho_{\rm GGE} 
= 
\frac{1}{Z} 
\exp 
\left( 
- \sum_{\sigma =\pm} \int_{0}^{\infty} \mathrm{d}\alpha~ \beta^\sigma(\alpha) Q^\sigma (\alpha) 
\right)  \,,
\label{rhoGGEdiag}
\ee 
where the factor $Z$ is chosen such as to ensure the normalization $\mathrm{Tr}\varrho_{\rm GGE} =1 $.

\subsubsection{Non-diagonal scattering}

\label{sec:miloszlimitnondiag}

\begin{figure}
\begin{center}
\begin{tikzpicture}[scale=1]
\begin{scope}[shift={(-4,2.)}]
\filldraw[blue!5] (-0.75,-0.08) --(0.75,-0.08) -- (0,-2)  -- cycle;
\fill [blue!40] (-0.75,-0.08) rectangle (0.75,0.4);
\draw[<->] (1.65,0.15) -- node[above]{$\delta$} (1.95,0.15);
\draw [decoration={brace,raise=0.25}, decorate
] (-0.75,0.5) -- node[above]{$n$} (0.75,0.5);
\foreach \x in {-0.75,-0.45,...,0} { \draw[fill=blue!100] (\x,0) circle (0.08);\draw[fill=blue!100] (-\x,0) circle (0.08);}
\foreach \x in {-1.05,-1.35,...,-3} { \draw[fill=blue!0] (\x,0) circle (0.08); \draw[fill=blue!0] (-\x,0) circle (0.08);}
\draw[thick,gray] (-2.9,-2) -- (2.9,-2);
\draw (-0.75,-0.18) -- (-0.05,-1.9);
\draw (0.75,-0.18) -- (0.05,-1.9);
\fill[blue!100] (0,-2) circle (0.08);
\node[text width=1.5cm] at (-1.75,-1) {$\delta \to 0$ \\ $n$ fixed};
\end{scope}

\begin{scope}[shift={(-4,-2.)}]
\filldraw[blue!5] (-0.75,-0.08) --(0.75,-0.08) -- (0.3,-2) -- (-0.3,-2) -- cycle;
\fill [blue!40] (-0.75,-0.08) rectangle (0.75,0.4);
\draw[<->] (1.65,0.15) -- node[above]{$\delta$} (1.95,0.15);
\draw [decoration={brace,raise=0.25}, decorate
] (-0.75,0.5) -- node[above]{$n$} (0.75,0.5);
\foreach \x in {-0.75,-0.45,...,0} { \draw[fill=blue!100] (\x,0) circle (0.08);\draw[fill=blue!100] (-\x,0) circle (0.08);}
\foreach \x in {-1.05,-1.35,...,-3} { \draw[fill=blue!0] (\x,0) circle (0.08); \draw[fill=blue!0] (-\x,0) circle (0.08);}
\draw[thick,gray] (-2.9,-2) -- (2.9,-2);
\draw (-0.75,-0.18) -- (-0.3,-1.9);
\draw (0.75,-0.18) -- (0.3,-1.9);
\filldraw[blue!100] (-0.3,-2) -- (0.3,-2) -- (0.3,-1.9)-- (-0.3,-1.9) -- cycle;
\draw [<->] (-0.3,-2.2) -- node[below] {$\alpha$} (0.3,-2.2);
\node[text width=1.5cm] at (-1.75,-1) {$\delta \to 0$ \\ $n \delta$ fixed};
\end{scope}

\begin{scope}[shift={(4,2.)}]
\filldraw[blue!5] (-0.75,-0.08) --(0.75,-0.08) -- (0,-2)  -- cycle;
\filldraw[blue!45, rounded corners = 8 pt] (-2.9,-0.08) --(2.9,-0.08) -- (2.5,-0.05) -- (2,0.) -- (1,0.15) -- (0.5,0.35) -- (0,0.5) -- (-0.5,0.35) -- (-1,0.15) -- (-2,0) -- (-2.5,-0.05) -- cycle;
\foreach \x in {-0.45,-0.15,...,0} { \draw[fill=blue!100] (\x,0) circle (0.08);\draw[fill=blue!100] (-\x,0) circle (0.08);}
\draw[<->] (1.65,0.15) -- node[above]{$\delta$} (1.95,0.15);
\draw [decoration={brace,raise=0.25}, decorate
] (-0.75,0.5) -- node[above]{$n$} (0.75,0.5);
\foreach \x in {-0.75,-0.45} { \draw[fill=blue!75] (\x,0) circle (0.08);\draw[fill=blue!75] (-\x,0) circle (0.08);}
\foreach \x in {-1.05,-1.35,-1.65} { \draw[fill=blue!50] (\x,0) circle (0.08);\draw[fill=blue!50] (-\x,0) circle (0.08);}
\foreach \x in {-1.65,-1.95,...,-3} { \draw[fill=blue!20] (\x,0) circle (0.08); \draw[fill=blue!20] (-\x,0) circle (0.08);}
\draw[thick,gray] (-2.9,-2) -- (2.9,-2);
\draw (-0.75,-0.18) -- (-0.05,-1.9);
\draw (0.75,-0.18) -- (0.05,-1.9);
\fill[blue!100] (0,-2) circle (0.08);
\node[text width=1.5cm] at (-1.75,-1) {$\delta \to 0$ \\ $n$ fixed};
\end{scope}

\begin{scope}[shift={(4,-2.)}]
\filldraw[blue!5] (-0.75,-0.08) --(0.75,-0.08) -- (0.3,-2) -- (-0.3,-2) -- cycle;
\filldraw[blue!45, rounded corners = 8 pt] (-2.9,-0.08) --(2.9,-0.08) -- (2.5,-0.05) -- (2,0.) -- (1,0.15) -- (0.5,0.35) -- (0,0.5) -- (-0.5,0.35) -- (-1,0.15) -- (-2,0) -- (-2.5,-0.05) -- cycle;
\foreach \x in {-0.45,-0.15,...,0} { \draw[fill=blue!100] (\x,0) circle (0.08);\draw[fill=blue!100] (-\x,0) circle (0.08);}
\draw[<->] (1.65,0.15) -- node[above]{$\delta$} (1.95,0.15);
\draw [decoration={brace,raise=0.25}, decorate
] (-0.75,0.5) -- node[above]{$n$} (0.75,0.5);
\foreach \x in {-0.75,-0.45} { \draw[fill=blue!75] (\x,0) circle (0.08);\draw[fill=blue!75] (-\x,0) circle (0.08);}
\foreach \x in {-1.05,-1.35,-1.65} { \draw[fill=blue!50] (\x,0) circle (0.08);\draw[fill=blue!50] (-\x,0) circle (0.08);}
\foreach \x in {-1.65,-1.95,...,-3} { \draw[fill=blue!20] (\x,0) circle (0.08); \draw[fill=blue!20] (-\x,0) circle (0.08);}
\draw[thick,gray] (-2.9,-2) -- (2.9,-2);
\draw (-0.75,-0.18) -- (-0.3,-1.9);
\draw (0.75,-0.18) -- (0.3,-1.9);
\draw [<->] (-0.3,-2.2) -- node[below] {$\alpha$} (0.3,-2.2);
\node[text width=1.5cm] at (-1.75,-1) {$\delta \to 0$ \\ $n \delta$ fixed};
\filldraw[blue!100, rounded corners=5pt] (-2,-2) -- (2,-2) -- (1.5,-1.99)-- (1,-1.97) -- (0,-1.8) -- (-1,-1.97) -- (-1.5,-1.99) -- cycle;
\end{scope}
\end{tikzpicture} 
\end{center}
\caption{
Procedure detailed in section \ref{sec:miloszlimitnondiag} for taking the scaling limit of local (left panels) and quasilocal (righ panels) lattice charges. On all of the four panels the lattice is represented at the top, and the field theory limit at the bottom. In the quasilocal case, the charge densities have an exponentially decaying norm, but a typical range $ \sum_{r=2}^{\infty} r ||q_{j,n}^{\pm [r]}||^2 $ increasing with the derivation order $n$, and which we have simply indicated as $n$ itself on the figure.  
}
\label{fig:limitingprocedure}
\end{figure}
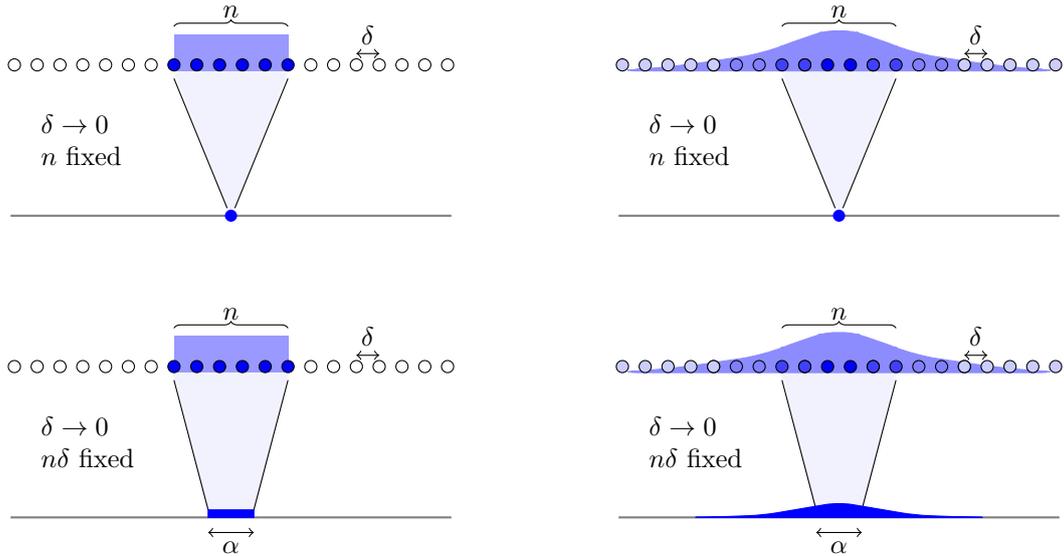

We now need to generalize the limiting procedure reviewed above in order to reveal all the conserved charges that are necessary to fix particle and string densities in sine-Gordon. For the densities of physical particles (solitons, antisolitons), the limiting procedure looks exactly the same as that of \cite{milosz}. One considers the ultra local conserved charges generated by $X_1(\alpha)$, and takes the two different continuum limits as described in the previous paragraph. In the first case, depicted on the top-left panel of figure \ref{fig:limitingprocedure}, this produces the known integer spin, local charges of sine-Gordon. In the second case, bottom-left panel, it produces a continuum of semilocal charges with support on a finite region of space.

The more interesting new problem we now face is to describe the two continuum limiting procedures starting from the quasilocal lattice charges, generated by $X_j(\alpha)$ with $j>1$. To understand the analogue of the procedures of \cite{milosz} for the quasilocal lattice charges, we need to understand for a given value of $j$ how the support of the conserved charges $Q_{j,n}^{\pm}$ is affected by changing the value of the derivation order $n$. In contrast to the local case $j=1$, the quasilocal ones have support that extends through all space, but with an exponentially decaying Hilbert-Schmidt norm, $||q_{j,n}^{\pm [r]}||^2_{\rm HS}$. Knowing this norm, we can study what is the ``typical range", $ \sum_{r=2}^{\infty} r ||q_{j,n}^{\pm [r]}||^2 $, which tells us how the corresponding profile of Hilbert-Schmidt norms distribution widens as we increase values of $n$. 
One particular important question we need to answer is, does the typical length of the Hilbert-Schmidt norm, $||q_{j,n}^{\pm [r]}||^2_{\rm HS}$,  increase linearly with $n$, such that the limiting procedure of \cite{milosz} can be generalized without major modifications?

This question is difficult to answer in general for all the charges, $Q_{j,n}^{\pm }$, however some qualitative idea can be achieved by  computing explicitly the Hilbert-Schmidt norm for increasing values of $n$ as a function of $r$ (see Figure \ref{fig:qnorms}).  For the local charges (with $j=1$) it can be seen that the Hilbert-Schmidt norm is non-zero only for $r$ smaller than a given maximum value proportional to $n$, which means that the charges have a finite support which increases linearly with $n$. For the quasilocal charges (where we show the plots for $j=2$), the norms are nonzero for all values of $r$, however they have the same exponential decay, as is proven in Appendix \ref{app:quasilocal}. Interestingly, it is very clear that the norm distribution function becomes more extended as $n$ increases, such that the typical length indeed does increase with $n$. 
The set of accessible values of $n$ makes very difficult to conjecture anything about the nature of the growth, however guided by intuition of the local case we will assume in the following that it is linear in this case too.

We now return to the issue of taking the continuum limit of these charges, with the different procedures of \cite{milosz}, for a given value of $j$. The first limit consists on fixing a finite value of $n$, and taking the continuum limit $\delta\to0$, such that $n \delta\to0$. In this finite $n$ scenario, the exponential curves shown in Figure \ref{fig:qnorms} become narrower as we reduce the lattice spacing, leading in the continuum to a vanishing typical length of order $n \delta$. The vanishing of the typical length implies that these charges in the continuum limit are in fact completely local. This is illustrated on the top-right panel of figure \ref{fig:limitingprocedure}. As we have seen in the previous section, this continuum limit produces a discrete set of charges with fractional (and coupling-constant dependent) Lorentz spins given by $s=2k\gamma/(\pi-\gamma)$, for integer values of $k$. 

The second continuum limit from \cite{milosz} consists in considering charges with infinitely large values of $n$, such that $n \delta$ is fixed to be finite as $\delta\to 0$. In this limit, which we depict on the bottom-right panel of figure \ref{fig:limitingprocedure}, the effect of reducing $\delta$ is to squeeze the spatial width of the Hilbert-Schmidt norm, while increasing $n$ has exactly the opposite effect. This procedure yields a continuum of semilocal conserved charges $Q_{j}^\pm(\alpha)$ in the field theory, with an exponentially decaying Hilbert-Schmidt norm over space, whose typical length increases smoothly with the continuous parameter $\alpha = n \delta$.
%
%

%
Resultingly, the density matrix (\ref{rhoGGEdiag}) obtained in \cite{milosz} in the case of diagonal theories should generalize in the non-diagonal case under the form 
\be 
\varrho_{\rm GGE} 
= 
\frac{1}{Z} 
\exp 
\left( 
- \sum_{j}\sum_{\sigma =\pm}\int_{0}^{\infty} \mathrm{d}\alpha~ \beta_j^\sigma(\alpha) Q_{j}^\sigma (\alpha) 
\right)  \,, 
\label{rhoGGEnondiag}
\ee 
where the index $j$ is summed over the coupling-dependent set of existing magnons (see sections above), and where the factor $Z$ is chosen such as to ensure the normalization $\mathrm{Tr}\varrho_{\rm GGE} =1 $.

\section{Conclusion}

We have established the existence of a set of conserved charges in the sine-Gordon model which are local, and have fractional Lorentz spin. These were obtained starting from the observation that sine-Gordon arises as the continuum limit of a spatially inhomogeneous version of the spin-$\frac{1}{2}$ XXZ chain. Using the algebraic Bethe ansatz formalism, we have shown that the inhomogeneous spin chain also has a set of quasilocal conserved charges, as has been previously established for the standard homogeneous case. The new conserved charges in the sine-Gordon field theory are then obtained by a careful continuum limiting procedure (where the quasilocal lattice charges become local, fractional spin field theory charges).
Let us point out that while we have precisely worked out the expression of the charges only for integer values of the parameter $p$, related to the sine-Gordon coupling by equation (\ref{eq:parameterp}), similiar results may easily be found for any rational value $p>1$, upon changing accordingly the string content of the theory. Since such values cover densely the sine-Gordon repulsive regime, our conclusions may be extended to the latter. Extension to the attractive regime ($p \leq 1$) requires some more care, but should be feasible. In particular, quasilocal charges may be similarly exist on the light-cone lattice, as they do in the homogeneous XXZ case \cite{Ilievski-stringcharge}. Because the structure of eigenstate is in this case slightly more complicated, due to the existence of breathers, the procedure followed in this work for deriving the corresponding eigenvalues of these charges requires some adaptation. 

The existence of these local charges has very practical consequences in the context of quantum quenches and equilibration. These charges have been shown to play a crucial role in describing the long-time stationary state after a quench, and need to be included in the GGE description. The use of these charges is to fix the density distributions of auxiliary particles in non-diagonal scattering theories, such as magnons and strings. This is because the usual integer spin conserved charges of integrable field theory can only measure kinematic properties of physical particles, but are not able to fix any information about the auxiliary particles. To completely fix the density distributions of auxiliary particles in stationary states with the GGE formalism, one needs to consider different procedures for taking the continuum limit of the quasilocal lattice charges. In this context, we have shown that the limiting procedure proposed in \cite{milosz} for the local lattice charges, can also be used for the new quasilocal lattice charges.

While the present construction has been concerned only with one model, the sine-Gordon model, we believe that it should go similarly for other theories with non-diagonal scattering. Starting from an appropriate lattice discretization, one should in principle be able to build (quasi)local conserved charges in correspondence with the various types of stable (physical or magnonic) particle species. These should be, in turn, the building block of a complete GGE density matrix.  Some interesting examples include the Bullough-Dodd model with imaginary coupling and the O(3) nonlinear sigma model, but  all require an important amount of work. In particular, while the former may be obtained as the continuum limit of the so-called Izergin-Korepin integrable spin chain \cite{a22}, the existence of associated lattice quasilocal conserved charges has not yet been established.

The charges found in our paper might have a deep mathematical connection with other conserved charges of sine-Gordon that have been previously discovered. In particular, the fractional spins of our new conserved charges matches the spin of the non-local charges that have been found in \cite{Bernard1,Bernard2,Russians,Russians2}. 
The greatest difference between our new work and previously known charges is that we have shown, starting from the lattice discretization, that our conserved charges are local in the scaling limit, while the other previously known charges are, at least manifestly, non-local. 
A likely scenario is that our charges could form some Abelian subset of the quantum group charges of Bernard and LeClair \cite{Bernard1,Bernard2}, and have some relation, possibly of the linear form, with the commuting charges of Bazhanov {\it et.al} \cite{Russians,Russians2}. If this were the case, it would indicate that the latter, despite their manifestly non-local expression in terms of vertex operators (\ref{chargesBLZ}), are in fact local operators. Elucidating the relation between our charges and those of \cite{Bernard1,Bernard2,Russians,Russians2} is however a highly non-trivial task, which we leave for future work.

There are many applications of the new sine-Gordon charges to explore in the future. For instance, we would like to use these to find the exact GGE descriptions of stationary states corresponding to different initial states (for example, the simple initial states considered in \cite{Bertini,marton}). An important task in this case is to evaluate the expectation values of these new charges on the considered initial state. Since our charges are most easily formulated on an eigenbasis of particles and magnons densities rather than on asymptotic states made of particles with individually fixed topological charge, this clearly requires some work. Our charges and the currents associated with them can also be used to generalize the recently proposed  integrable hydrodynamics description of transport phenomena \cite{hydro} for non-diagonal scattering theories. Finally, the construction of additional charges in sine-Gordon following our procedure seems to be adaptable to the further set of ``non-unitary'' charges, which in the XXZ case have found applications in the study of spin transport \cite{prosen-11,prosen-13,ppsa-14,pv-16}.  It would be  interesting to understand the meaning of these charges in the sine-Gordon model, as well as possible applications to its transport properties \cite{doyonTBA}.

\subsection*{Acknowledgments}

We acknowledge enlightening discussions with Toma\v{z} Prosen, Denis Bernard, M\'{a}rton Kormos and in particular Hubert Saleur and G\'{a}bor Tak\'{a}cs, as well as Pasquale Calabrese and Lorenzo Piroli for useful comments on the manuscript.  

 This work has been supported by the ERC under Starting Grant 279391 EDEQS.

\section*{Appendices}

\begin{appendix}

\section{Bethe equations for the XXZ spin chain / six-vertex model}
\label{sec:XXZ}

In this appendix we review the Bethe ansatz solution of the XXZ quantum spin chain and the associated six-vertex statistical mechanics model.
The XXZ chain is defined on a chain of $N$ spins 1/2 with periodic boundary conditions, and the Hamiltonian is written in terms of the local Pauli matrices as
\be
H = \frac{J}{4}\sum_{i=1}^{N} \left[ \sigma^x_i  \sigma^x_{i+1} + \sigma^y_i  \sigma^y_{i+1} + \Delta (\sigma^z_i  \sigma^z_{i+1}-1) \right] \,.
\label{eq:HXXZ}
\ee
In the following we restrict to values of the real anisotropy parameter corresponding to the gapless regime, namely  $|\Delta| \leq 1$, and use the parametrization $\Delta = \cos\frac{\pi}{\tilde{p}+1}$.

The underlying two-dimensional statistical mechanics model is the six-vertex model, for which one defines the family of mutually commuting row-to-row transfer matrices
\be 
T(u | \{u_i \}_{i=1, \ldots N}) = \mathrm{tr}_0\left(R_{0,N}(u- u_N)\ldots R_{0,2}(u- u_2)R_{0,1}(u- u_1) \right) \,.
\label{eq:TXXZ}
\ee
The $R$ matrix was introduced in equation (\ref{Rmatrix}) (the parameter $\gamma$ used in ((\ref{Rmatrix})) should here be taken as $\gamma = \frac{\pi}{\tilde{p}+1}$), an the parameters $\{u_i \}_{i=1,\ldots N}$, which may be generically complex numbers, encode possible inhomogeneities. The Hamiltonian (\ref{eq:HXXZ}) is recovered from the homogeneous case  $\{u_i =0\}$ as 
\be
H = \left. \frac{\mathrm{d}}{\mathrm{d}u}  \log T(u | \{0\}_{i=1, \ldots N})  \right|_{u=0}    \,.
\ee

The Bethe ansatz construction \cite{korepin} allows to express the eigenstates of the transfer matrix (\ref{eq:TXXZ}) in terms of a set of quasi momenta $\{\lambda_k\}$ (the so called Bethe roots), solution of the Bethe equations 
\be
\prod_{j=1}^N
\frac{\sinh \left(\frac{1}{\tilde{p}+1} (\lambda_k + i \pi /2) - i u_j \right)}{\sinh \left(\frac{1}{\tilde{p}+1} (\lambda_k - i \pi /2) - i u_j \right)}
= 
\prod_{l (\neq k)}
\frac{\sinh \left( \frac{1}{\tilde{p}+1}(\lambda_k - \lambda_l + i \pi ) \right)}{\sinh \left( \frac{1}{\tilde{p}+1} (\lambda_k - \lambda_l - i \pi ) \right) }, 
\label{eq:bethe_eq}
\ee
In the homogeneous case, the solutions of \eqref{eq:bethe_eq} were shown \cite{Takahashi} to organize themselves into regular patterns in the complex plane called ``strings". It is easy to see however that the arguments that lead to this classification also hold in the case where the inhomogeneous parameters are purely imaginary, to which we will restrict in the following.  
Following \cite{Takahashi}, a $m$-string corresponds to a set of Bethe roots parametrized as
\be
\lambda_{k}^{\nu,(m)}=  \lambda^{(m)}_{k} +  {i \pi} \left(\nu-\frac{m+1}{2}\right)+\delta^{j,(m)}_{k}\,\qquad \nu=1,\ldots, m  \,,
\label{string_hp}
\nonumber
\ee
where $\lambda^{j,(m)}_{k}$ is a real number called the string center, and the numbers $\delta^{j,(m)}_{k}$ are deviations from a perfect string which vanish exponentially with the system size and are therefore neglected in the so-called {\it string hypothesis} \cite{Takahashi}. In addition one may also encounter strings of odd parity, the so-called $(m-)$-strings, whose center $ \lambda^{(m-)}_{k}$ is shifted from the real axis by $i \frac{ \pi}{2}(\tilde{p}+1)$. The string content of the model, namely the allowed values of $m$ are fixed by the value of $\tilde{p}$, in particular we will specify here to the case where $\tilde{p}$ is some positive integer, and refer to \cite{Takahashi} for an exhaustive description.
In this case the set of allowed strings is \cite{Takahashi}
\bea
\mbox{ even parity : } m&=&1,2,\ldots \tilde{p} \,, \nonumber \\
\mbox{odd parity : }& &(1-) \,. \nonumber 
\eea

In the limit $L \to \infty$, the string centers become dense on the real axis, and the eigenstates are conveniently described by smooth distribution functions $\rho_m(\lambda)$ (one for each type of string), as well as hole distribution functions $\rho_{n}^h(\lambda)$ which are a generalization to the interacting case of the hole distributions of an ideal Fermi gas at finite temperature \cite{korepin,Takahashi}. 
The Bethe equations can be recast in the following linear form for densities 
\be
\rho_m(\lambda) + \rho_m^h(\lambda) =  \frac{1}{N} \sum_{j=1}^N a_m(\lambda- \pi v_j/\gamma) - \left(\sum_{n} a_{m,n} \star \rho_n\right)(\lambda) \qquad \qquad (m= 1,\ldots, \tilde{p},(1-)) \,,
\label{eq:BetheTakahashi} 
\ee
where we have introduced the real inhomogeneous parameters $\{u_j\}_{j=1, \ldots N}$ as $u_j = i v_j$, the symbol $\star$ denotes a convolution,  and the different kernels are given by \cite{Takahashi}
\bea
a_m(\lambda) &=& \phi_{\frac{m}{2}}(\lambda)   ~~~~~~~~~~~~ (m \leq p)  \nonumber \\
a_{(1-)}(\lambda) &=& \phi_{\frac{1}{2}}(\lambda+ i \pi/2)  \nonumber    \\
a_{m,n}(\lambda) &=& (1-\delta_{m,n}) \phi_{\frac{|m-n|}{2}}(\lambda)+2\phi_{\frac{|m-n|}{2}+1}(\lambda)+\ldots+2\phi_{\frac{m+n}{2}-1}(u)+\phi_{\frac{m+n}{2}}(\lambda) ~ (m,n \leq p\,,m\neq n)  \nonumber  \\
a_{m,(1-)}(\lambda) &=& a_{(1-),m}(\lambda) = 2 \phi_{{m-1 \over 2}}(\lambda+i \pi/2) +2 \phi_{{m+1 \over 2}}(\lambda+i \pi/2)~~~~~~~   (m \leq p)      \nonumber  \\
a_{(1-),(1-)}(\lambda) &=& \phi_{1}(\lambda) \,,
\label{eq:BAEkernels}
\eea
with
\be
\phi_k(\lambda) =\frac{\mathrm{d}}{\mathrm{d}\lambda} \frac{i}{2}\log\frac{\sinh \left(\frac{1}{\tilde{p}+1} (\lambda - i k \pi) \right)}{\sinh \left(\frac{1}{\tilde{p}+1} (\lambda + i k \pi) \right)} \,.
\label{eq:phidef}
\ee
For reference in the main text we introduce the corresponding Fourier transforms. The convention we use is 
\be \widehat{f}(\omega) = \int_{-\infty}^{\infty} \mathrm{d}\lambda e^{i \frac{\lambda}{\tilde{p}+1} \omega} f(\lambda) \,.  
\ee 
The transform of the various functions used above are then given by
\bea
\widehat{\phi}_k(\omega) &=& \frac{\sinh \left( \omega \left( \frac{\pi}{2}-k\frac{\pi}{\tilde{p}+1} \right) \right)}{\sinh\frac{\omega \pi}{2}}  \qquad \mbox{(for $0<k < \frac{\tilde{p}+1}{2}$)\,,}  \nonumber \\
\widehat{a}_m(\omega) &=&\frac{\sinh \omega\left(  \frac{\pi}{2} - \frac{m}{2} \frac{\pi}{\tilde{p}+1} \right)}{ \sinh \frac{\omega \pi}{2}}  \nonumber \\
\widehat{a}_{(1-)}(\omega) &=& 
-\frac{\sinh \left(\frac{\omega}{2}\frac{\pi}{\tilde{p}+1} \right)}{ \sinh \frac{\omega \pi}{2}}
 \nonumber    \\
\widehat{a}_{m,n}(\omega) &=& 2 \coth\left(\frac{\omega }{2}\frac{\pi}{\tilde{p}+1}\right)
\frac{ \sinh \left(\frac{\omega  m }{2}\frac{\pi}{\tilde{p}+1}\right)  \sinh \omega\left(  \frac{\pi}{2} - \frac{n }{2}\frac{\pi}{\tilde{p}+1} \right) }{ \sinh \frac{\omega \pi}{2}} - \delta_{m,n} ~~~~(m\leq n \leq \tilde{p})  \nonumber  \\
\widehat{a}_{m,(1-)}(\omega) = \widehat{a}_{(1-),m}(\omega)&=&  -2 
\frac{ \sinh \left(\frac{\omega  m }{2}\frac{\pi}{\tilde{p}+1}\right)  \sinh \left(\frac{\omega   }{2}\frac{\pi}{\tilde{p}+1}\right)  }{ \cosh \frac{\omega \pi}{2}} - \delta_{m,p} 
~~~~~~~   (m \leq \tilde{p})      \nonumber  \\
\widehat{a}_{(1-),(1-)}(\omega)  &=&
\frac{\sinh \omega\left(  \frac{\pi}{2} - \frac{\pi}{\tilde{p}+1} \right)}{ \sinh \frac{\omega \pi}{2}} 
 \,.
 \label{eq:BAEFourierkernels}
\eea 


\section{Proof of quasilocality}
\label{app:quasilocal}

In this appendix we prove that the light-cone lattice operators (\ref{eq:Xjstagdef}), or rather a suitable $\gamma$-dependent subset thereof (described in the main text), are quasilocal.
First, it is convenient to introduce the operators
\bea
\widehat{X}_j (\alpha ) &=&
 \frac{ T_{j}\left(\frac{\gamma}{\pi} i \alpha -  \gamma \right)\partial_\alpha T_{j} \left( \frac{\gamma}{\pi} i\alpha   \right)   }{\left[\epsilon_j(i \alpha  )\right]^N}  
\,,
\label{eq:defXj}
\eea
where
\bea
\epsilon_{j}(u) &=& 
\sin \left(u + \gamma\frac{j+1}{2} \right) \sin \left(u - \gamma\frac{j+1}{2} \right) \,.
\label{eq:epsilon12}
\eea
In the large $N$ limit these can be related to ${X}_j$ through the so-called inversion relation (\ref{eq:inversionrelation}) \cite{fe2-13,Ilievski-stringcharge}, which holds in the physical domain (\ref{eq:physicaldomain}) as long as $j$ belongs to the suitably chosen set. Namely
\be
\frac{X_j(\alpha)}{N}
\simeq 
\frac{\widehat{X}_j(\alpha)}{N}
+  \partial_{\alpha} \log   \sin \left( \frac{\gamma}{\pi} i \alpha + \frac{j \gamma}{2} \right)  
\,,
\label{eq:aux_2}
\ee
so the quasilocality of $X_j$ can be deduced from that of $\widehat{X}_j$.
In order to write the latter as a sum of densities with increasing support, it is convenient to introduce, as done for instance in \cite{imp-15,impz-16}, the decomposition of the Lax operators $L^{(j)}_i$ over a basis of Pauli matrices
\be
L^{(j)}_i(u) = \sum_{\alpha = 0,z,\pm} \mathcal{A}_\alpha(u)  \sigma_i^\alpha  \,,
\nonumber
\ee
where $\mathcal{A}_0(u),\mathcal{A}_z(u),\mathcal{A}_+(u),\mathcal{A}_-(u)$ are operators acting on the auxilliary spin-$j/2$ representation, for which we omit the superscript ${}^{(j)}$ for simplicity. More explicitly, the action of these operators is defined in terms of the spin-$j/2$ generators $S^{(j)z}, S^{(j)+}, S^{(j)-}$ as 
\bea
\mathcal{A}_0(u) &=&  \sin\left(u + \frac{\gamma}{2}\right) ~\cos (\gamma S^{(j)z} ) \nonumber \\
\mathcal{A}_z(u) &=&  \cos\left(u + \frac{\gamma}{2}\right) ~\sin (\gamma S^{(j)z} ) \nonumber \\
\mathcal{A}_+(u) &=&  \sin \gamma ~ S^{(j)-} \nonumber \\
\mathcal{A}_-(u) &=&  \sin \gamma ~ S^{(j)+}  \,.
\nonumber
\eea
From there $\widehat{X}_j(\alpha )$ can be decomposed as 
\bea
\widehat{X}_j(\alpha ) &=& \frac{i }{\left[\epsilon_j(i \alpha)\right]^N} 
\partial_{v}\left(
\sum_{\substack{\alpha_1,\ldots,\alpha_N   \\ \beta_1,\ldots,\beta_N } }
\mathrm{tr} \left[  \substack{ \mathcal{A}_{\alpha_N}(u+ i \Theta/2) \\ \mathcal{A}_{\beta_N}(v + i \Theta/2)} \ldots  \substack{ \mathcal{A}_{\alpha_1}(u- i \Theta/2) \\ \mathcal{A}_{\beta_1}(v - i \Theta/2)} \right]
(\sigma_1^{\alpha_1} \sigma_1^{\beta_1}) \ldots (\sigma_N^{\alpha_N} \sigma_N^{\beta_N})
\right)_{\substack{ u= i \alpha - \gamma \\  v = i \alpha}} \,
\nonumber
\\
 &=& \frac{i }{\left[\epsilon_j(i \alpha)\right]^N} 
\partial_{v}\left(
\sum_{\alpha_1,\ldots,\alpha_N  }
\mathrm{tr} \left[
\mathcal{B}_{\alpha_N}\left(\substack{u+ i\Theta/2 \\ v + i \Theta/2 } \right)
\ldots
\mathcal{B}_{\alpha_1}\left(\substack{u-i \Theta/2 \\ v - i\Theta/2 } \right)
 \right]
\sigma_1^{\alpha_1}  \ldots \sigma_N^{\alpha_N}
\right)_{\substack{ u= i \alpha - \gamma \\  v = i \alpha}} \,,
\label{eq:XjB}
\eea
where the trace is now over the product of two spin-$j/2$ auxilliary spaces and the notation  $\substack{ \mathcal{A}_{\alpha_i} \\ \mathcal{A}_{\beta_i} }$ has been used to describe the tensor product of two $\mathcal{A}$ acting on each tensor respectively. We also recall that the rescaled staggering parameter $\Theta$ has been introduced in the main text, see equation (\ref{eq:tiltedLambda}).
In the second line of equation (\ref{eq:XjB}) we have used the multiplication properties of Pauli matrices and introduced the operators
\bea
\mathcal{B}_0 \left(\substack{u \\v} \right) &=&  
 \substack{ \mathcal{A}_{0}(u) \\ \mathcal{A}_{0}( u)}
+
 \substack{ \mathcal{A}_{z}(u) \\ \mathcal{A}_{z}( v)}
+
\frac{1}{2}
 \substack{ \mathcal{A}_{+}(u) \\ \mathcal{A}_{-}( v)}
+
\frac{1}{2}
 \substack{ \mathcal{A}_{-}(u) \\ \mathcal{A}_{+}(v)}
 \\
\mathcal{B}_z\left(\substack{u \\v} \right)  &=& 
 \substack{ \mathcal{A}_{z}(u) \\ \mathcal{A}_{0}( v)}
+
 \substack{ \mathcal{A}_{0}(u) \\ \mathcal{A}_{z}(v)}
+
\frac{1}{2}
 \substack{ \mathcal{A}_{+}(u) \\ \mathcal{A}_{-}( v)}
-
\frac{1}{2}
 \substack{ \mathcal{A}_{-}(u) \\ \mathcal{A}_{+}(v)}
 \\
\mathcal{B}_+\left(\substack{u \\v} \right)  &=& 
 \substack{ \mathcal{A}_{0}(u) \\ \mathcal{A}_{+}(v)}
+
 \substack{ \mathcal{A}_{+}(u) \\ \mathcal{A}_{0}( v)}
+
 \substack{ \mathcal{A}_{z}(u) \\ \mathcal{A}_{+}( v)}
-
 \substack{ \mathcal{A}_{+}(u) \\ \mathcal{A}_{z}(v)}
 \\
\mathcal{B}_-\left(\substack{u \\v} \right)  &=&  
 \substack{ \mathcal{A}_{0}(u) \\ \mathcal{A}_{-}( v)}
+
 \substack{ \mathcal{A}_{-}(u) \\ \mathcal{A}_{0}( v)}
-
 \substack{ \mathcal{A}_{z}(u) \\ \mathcal{A}_{-}( v)}
+
 \substack{ \mathcal{A}_{-}(u) \\ \mathcal{A}_{z}(v)}
\eea
acting on the tensor product. Introducing further 
\bea 
\widetilde{\mathcal{B}}_{0,+,-,z}(\alpha)
&=&
\frac{1 }{\epsilon_j(i \alpha)}  \mathcal{B}_{0,+,-,z}\left(\substack{i \alpha - \gamma \\i \alpha} \right) \nonumber \\
\widetilde{\mathcal{B}}^\partial_{0,+,-,z}(\alpha)
&=&
\frac{i }{\epsilon_j(i \alpha)}  \left.\partial_v\mathcal{B}_{0,+,-,z}\left(\substack{u \\v} \right)  \right|_{\substack{ u= i \alpha - \gamma \\  v = i \alpha}} \,,
\eea 
and using the cyclicity of the trace, we can recast (\ref{eq:XjB}) as
\bea
\widehat{X}_j(\alpha )  &=&  
\sum_{\substack{k=0 \\ k~{\rm odd}}}^{N} 
\sum_{\alpha_1, \ldots \alpha_N} \mathrm{tr} \left( 
\widetilde{B}^\partial_{\alpha_1}(\alpha + \Theta/2) \widetilde{B}_{\alpha_2}(\alpha - \Theta/2) \widetilde{B}_{\alpha_3}(\alpha + \Theta/2) \ldots \widetilde{B}_{\alpha_N}(\alpha - \Theta/2)
\right)
\sigma_k^{\alpha_1} \ldots\sigma_{k+N-1}^{\alpha_N} 
\nonumber \\
& & 
+
\sum_{\substack{k=0 \\ k~{\rm even}}}^{N} 
\sum_{\alpha_1, \ldots \alpha_N} \mathrm{tr} \left( 
\widetilde{B}^\partial_{\alpha_1}(\alpha - \Theta/2) \widetilde{B}_{\alpha_2}(\alpha + \Theta/2) \widetilde{B}_{\alpha_3}(\alpha -\Theta/2) \ldots \widetilde{B}_{\alpha_N}(\alpha +\Theta/2)
\right)
\sigma_k^{\alpha_1} \ldots\sigma_{k+N-1}^{\alpha_N} \,,
\nonumber 
\eea
where $\sigma_{k+N} \equiv \sigma_{k}$.
Therefore $\widehat{X}_j(\alpha ) $ can be written as a sum over densities $\widehat{x}_j(\alpha)$, namely
\be
\widehat{X}_j(\alpha) =  \sum_{k=0}^{N/2} \mathcal{P}_{2 k}\left( \widehat{x}_j(\alpha) \right) \,,
\nonumber
\ee
where $\mathcal{P}_{2k}$ represents a translation by $2k$ lattice sites to the right, and \bea
\widehat{x}_j(\alpha ) &=& 
\sum_{\alpha_1,\ldots,\alpha_N}
\mathrm{tr} \left[   
\widetilde{\mathcal{B}}^\partial_{\alpha_1} (\alpha + \Theta/2) 
\widetilde{\mathcal{B}}_{\alpha_2} (\alpha -\Theta/2)
\widetilde{\mathcal{B}}_{\alpha_3} (\alpha +\Theta/2)
\ldots 
\widetilde{\mathcal{B}}_{\alpha_N} (\alpha - \Theta/2)  
 \right]
 \sigma_1^{\alpha_1}  \ldots \sigma_N^{\alpha_N}  
 \nonumber \\
 &+& 
 \sum_{\alpha_1,\ldots,\alpha_N}
\mathrm{tr} \left[   
\widetilde{\mathcal{B}}^\partial_{\alpha_1} (\alpha -  \epsilon\Theta/2) 
\widetilde{\mathcal{B}}_{\alpha_2} (\alpha + \Theta/2)
\widetilde{\mathcal{B}}_{\alpha_3} (\alpha -\Theta/2)
\ldots 
\widetilde{\mathcal{B}}_{\alpha_N} (\alpha +\Theta/2)  
 \right]
 \sigma_1^{\alpha_2}  \ldots \sigma_N^{\alpha_N+1}  
 \,. \nonumber \\
\label{eq:xjdensities}
\eea

At this stage some known observations \cite{prosen-14} on the properties of the matrices $\widetilde{\mathcal{B}}_{\alpha}$ and $\widetilde{\mathcal{B}}^\partial_{\alpha}$ can be reproduced. In particular, it is seen that for any $\alpha$ the singlet left vector 
\be
\langle \psi_0 |  \equiv 
 (j+1)^{-1/2} \sum_{m=-j/2}^{j/2} (-1)^{j/2-m} \langle m | \otimes \langle-m | \,,
 \nonumber
\ee
is a left eigenvector of $\widetilde{\mathcal{B}}_0$ with eigenvalue $1$, and that for any real $\alpha$ all remaining eigenvalues are strictly smaller in absolute value. Further, it is checked that for all $\alpha$
\be
\langle \psi_0 |  \widetilde{\mathcal{B}}_{+,-,z}(\alpha) = 0 \,.
\label{eq:psi0ppty}
\ee
Inserting a resolution of the identity in the auxilliary space inside the trace in (\ref{eq:xjdensities}), one easily sees that the contributions from all eigenvectors different than $\langle \psi_0 | $ vanish exponentially with $N$, while equation (\ref{eq:psi0ppty}) allows to recast the $\langle \psi_0 | $  contribution as a sum of densities with increasing support, namely
\bea
\widehat{x}_j(\alpha ) &\stackrel{N \to \infty}{=}&
\sum_{\alpha_1,\ldots,\alpha_N   }
\langle \psi_0 |   
\widetilde{\mathcal{B}}^\partial_{\alpha_1} (\alpha + \Theta/2) 
\widetilde{\mathcal{B}}_{\alpha_2} (\alpha - \Theta/2)
\widetilde{\mathcal{B}}_{\alpha_3} (\alpha + \Theta/2)
\ldots 
\widetilde{\mathcal{B}}_{\alpha_N} (\alpha -\Theta/2)  | \psi_0 \rangle
 \sigma_1^{\alpha_1}  \ldots \sigma_N^{\alpha_N}  
 \nonumber \\
 & & 
 +
 \sum_{\alpha_1,\ldots,\alpha_N   }
\langle \psi_0 |   
\widetilde{\mathcal{B}}^\partial_{\alpha_1} (\alpha - \Theta/2) 
\widetilde{\mathcal{B}}_{\alpha_2} (\alpha + \Theta/2)
\widetilde{\mathcal{B}}_{\alpha_3} (\alpha - \Theta/2)
\ldots 
\widetilde{\mathcal{B}}_{\alpha_N} (\alpha+\Theta/2)  | \psi_0 \rangle
 \sigma_2^{\alpha_1}  \ldots \sigma_{N+1}^{\alpha_N}  
 \nonumber \\
&=& 
  \sum_{r=2}^{N}  \widehat{x}_j^{[r]}(\alpha)  \,,
\label{eq:xjdensitiesbis}
\eea
where $ \widehat{x}_j^{[r]}(\alpha) $ is a sum of two terms acting non trivially on $r$ consecutive sites only, namely 
\bea
 \widehat{x}_j^{[r]}(\alpha)   
&=& 
\sum_{\substack{
\alpha_1 , \ldots, \alpha_{r-1} \in \{0, +,-,z\} 
\\
\alpha_r  \in \{ +,-,z\}   
} } 
\langle \psi_0 |   \widetilde{\mathcal{B}}^\partial_{\alpha_1} (\alpha +\Theta/2) 
\widetilde{\mathcal{B}}_{\alpha_2} (\alpha - \Theta/2)
\ldots 
\widetilde{\mathcal{B}}_{\alpha_r} (\alpha -(-1)^r \Theta/2)  
| \psi_0 \rangle   
~
\sigma_1^{\alpha_1} \ldots
\sigma_r^{\alpha_r} 
\nonumber \\
& &+ 
\sum_{\substack{
\alpha_1 , \ldots, \alpha_{r-1} \in \{0, +,-,z\} 
\\
\alpha_r  \in \{ +,-,z\}   
} } 
\langle \psi_0 |   \widetilde{\mathcal{B}}^\partial_{\alpha_1} (\alpha -\Theta/2) 
\widetilde{\mathcal{B}}_{\alpha_2} (\alpha + \Theta/2)
\ldots 
\widetilde{\mathcal{B}}_{\alpha_r} (\alpha +(-1)^r \Theta/2)  
| \psi_0 \rangle   
~
\sigma_1^{\alpha_2} \ldots
\sigma_r^{\alpha_{r+1}} \,,
\nonumber \\
\label{eq:xjrdef}
\eea
and as identity on the rest of the chain.

As a temporary conclusion, we can therefore write similar expansions for the operators $X_j(\alpha)$ and the corresponding charges $Q_{j,n}^\pm$, namely
\bea
X_j(\alpha) &=&  \sum_{r=1}^{N} \sum_{k=0}^{N/2} \mathcal{P}_{2 k}\left(  x^{ [r]}_j(\alpha) \right)   \,, \nonumber \\ 
Q_{j,n}^\pm &=&   \sum_{r=1}^{N} \sum_{k=0}^{N/2} \mathcal{P}_{2 k}\left(   q^{\pm [r]}_{j,n} \right)   \,,
\eea
where
\bea 
x^{ [r]}_j(\alpha) &=& \widehat{x}^{ [r]}_j(\alpha ) \nonumber  \\
q^{\pm [r]}_{j,n} &=&  \left(\frac{\gamma}{\pi}\right)^n 
\left. \frac{\mathrm{d}^n}{\mathrm{d}\alpha^n}
\left(
 \widehat{x}^{ [r]}_j(\alpha + \Theta/2)  \pm(-1)^n  \widehat{x}^{ [r]}_j(\alpha - \Theta/2) \right) \right|_{\alpha=0}
 \label{eq:densitiesxjqj}
\eea

In order to establish the quasilocality of $X_j(\alpha)$ and $Q_{j,n}^\pm$, we must prove that the Hilbert-Schmidt norm of the corresponding finite-support densities decreases exponentially with $r$. For this, we compute the scalar products
\be 
\left \langle  \widehat{x}_j^{[r]}(\alpha) , \widehat{x}_j^{[r]}(\beta)  \right\rangle 
\equiv 
\frac{1}{2^N} \mathrm{Tr} 
\left(
\left( \widehat{x}_j^{[r]}(\alpha)\right)^\dag \widehat{x}_j^{[r]}(\beta) 
\right)
\nonumber
\ee 
Using (\ref{eq:xjrdef}), and the properties of Pauli matrices, the scalar products can be recast in terms of a set of matrices  acting now on the tensor product of four auxilliary spin-$j/2$ representations, namely 
\bea 
\left \langle  \widehat{x}_j^{[r]}(\alpha) , \widehat{x}_j^{[r]}(\beta)  \right\rangle 
&=&
\sum_{\epsilon=\pm 1}
\langle \substack{\psi_0  \\ \psi_0 } |
\mathcal{C}^{\partial \partial} 
\left( \substack{ \alpha + \epsilon \Theta/2 \\ \beta + \epsilon \Theta/2} \right)
\mathcal{C}\left( \substack{ \alpha - \epsilon \Theta/2 \\ \beta-  \epsilon \Theta/2} \right)
\ldots 
\mathcal{C} 
\left( \substack{ \alpha + (-1)^{r}\epsilon \Theta/2 \\ \beta +(-1)^{r} \epsilon \Theta/2} \right)
\mathcal{C}^{\rm right} 
\left( \substack{ \alpha + (-1)^{r+1}\epsilon \Theta/2 \\ \beta +(-1)^{r+1} \epsilon \Theta/2} \right)
| \substack{\psi_0  \\ \psi_0 } 
 \rangle  \,,
 \nonumber 
\eea
where, using once again transparent stacked notations for tensor products, 
\bea
\mathcal{C}^{\partial\partial}\left( \substack{ \alpha \\ \beta } \right) &=& 
\substack{   
\widetilde{\mathcal{B}}^\partial_0(\alpha)   \\ 
\widetilde{\mathcal{B}}^\partial_0 (\beta)^* 
}      
+
\substack{   
\widetilde{\mathcal{B}}^\partial_z(\alpha)   \\ 
\widetilde{\mathcal{B}}^\partial_z (\beta)^* 
}     
+
\frac{1}{2}
\substack{   
\widetilde{\mathcal{B}}^\partial_+(\alpha)   \\ 
\widetilde{\mathcal{B}}^\partial_+ (\beta)^* 
}    
+
\frac{1}{2}
\substack{   
\widetilde{\mathcal{B}}^\partial_-(\alpha)   \\ 
\widetilde{\mathcal{B}}^\partial_- (\beta)^* 
} 
\\ 
\mathcal{C}\left( \substack{ \alpha \\ \beta } \right) &=& 
\substack{   
\widetilde{\mathcal{B}}_0(\alpha)   \\ 
\widetilde{\mathcal{B}}_0 (\beta)^* 
}      
+
\substack{   
\widetilde{\mathcal{B}}_z(\alpha)   \\ 
\widetilde{\mathcal{B}}_z (\beta)^* 
}     
+
\frac{1}{2}
\substack{   
\widetilde{\mathcal{B}}_+(\alpha)   \\ 
\widetilde{\mathcal{B}}_+ (\beta)^* 
}    
+
\frac{1}{2}
\substack{   
\widetilde{\mathcal{B}}_-(\alpha)   \\ 
\widetilde{\mathcal{B}}_- (\beta)^* 
} 
\\ 
\mathcal{C}^{\rm right}\left( \substack{ \alpha \\ \beta } \right) &=& 
\substack{   
\widetilde{\mathcal{B}}_z(\alpha)   \\ 
\widetilde{\mathcal{B}}_z (\beta)^* 
}     
+
\frac{1}{2}
\substack{   
\widetilde{\mathcal{B}}_+(\alpha)   \\ 
\widetilde{\mathcal{B}}_+ (\beta)^* 
}    
+
\frac{1}{2}
\substack{   
\widetilde{\mathcal{B}}_-(\alpha)   \\ 
\widetilde{\mathcal{B}}_- (\beta)^* 
} 
 \,.
 \nonumber
\eea

We then check the following properties 
\begin{enumerate}
\item for any $\alpha, \beta$, $\langle \substack{\psi_0  \\ \psi_0 } |$ is a left eigenvector of ${\mathcal{C}}\left( \substack{ \alpha \\ \beta } \right)$ with eigenvalue $1$
\item $\langle \substack{\psi_0  \\ \psi_0 } | \mathcal{C}^{\rm right}\left( \substack{ \alpha \\ \beta } \right) =0$
\item for any real $\alpha$, the second leading eigenvalue $\tau_j(\alpha)$ of the product $\mathcal{C} \left( \substack{ \alpha + \Theta/2 \\ \alpha +\Theta/2 } \right)\mathcal{C} \left( \substack{ \alpha - \Theta/2 \\ \alpha - \Theta/2 } \right)$ is a common eigenvalue of the two operators $\mathcal{C} \left( \substack{ \alpha + \Theta/2 \\ \alpha +\Theta/2 } \right)$ and $\mathcal{C} \left( \substack{ \alpha - \Theta/2 \\ \alpha - \Theta/2 } \right)$, and is therefore of the form 
\be
\tau_j(\alpha) = f_j(\alpha-\Theta/2) f_j(\alpha+\Theta/2) \,,
\nonumber
\ee
where $f_j(\alpha)$ is found to have the form
\be
f_j(\alpha) = \frac{a(j,\gamma) + b(j,\gamma) \cosh 2\gamma\alpha/\pi + \frac{1}{2} \cosh(4\gamma\alpha/\pi)}{\left( \cos(j+1)\gamma  - \cosh(2\gamma\alpha/\pi) \right)^2} \,,
\label{eq:ajbj}
\ee
and $a(j,\gamma) $ and $b(j,\gamma)$  are some functions for which we could not obtain an analytical form. In any case, one has that $f_j(\alpha) \to 1$ as $\alpha \to \infty$.   
\item for any real $\alpha$, the second leading eigenvalue $\tau_j(\alpha)$ of the product $\mathcal{C} \left( \substack{ \alpha + \Theta/2 \\ -\alpha +\Theta/2 } \right)\mathcal{C} \left( \substack{ \alpha - \Theta/2 \\ -\alpha - \Theta/2 } \right)$ is strictly smaller than $\tau_{j}(\alpha)$. 
\end{enumerate}

From there, we are ready to conclude about the Hilbert-Schmidt norms of the densities (\ref{eq:densitiesxjqj}), namely
\bea
||x_j^{ [r]}(\alpha)||^2_{\rm HS} &=& \left \langle  (\widehat{x}_j^{ [r]}(\alpha))^\dag \widehat{x}_j^{[r]}(\alpha) \right\rangle 
\label{eq:xx} 
 \\
 ||q_{j,n}^{\pm[r]}||^2_{\rm HS} &=& 
 \left \langle  (q_{j,n}^{\pm[r]})^\dag q_{j,n}^{\pm[r]} \right\rangle 
\nonumber \\
 &=& 
 \frac{\mathrm{d}^n}{\mathrm{d}\alpha^n}\frac{\mathrm{d}^n}{\mathrm{d}\beta^n} 
 \left(
 \left \langle  (\widehat{x}_j^{\pm}(\alpha+\Theta/2))^\dag x_j^{[r]}(\alpha + \Theta/2) \right\rangle 
 +
 \left \langle  (\widehat{x}_j^{\pm}(\alpha-\Theta/2))^\dag x_j^{[r]}(\alpha - \Theta/2) \right\rangle 
 \right.
 \nonumber \\ 
 & & 
 \left.
  \left.
 \pm(-1)^n
 \left \langle  (\widehat{x}_j^{\pm}(\alpha+\Theta/2))^\dag x_j^{[r]}(\alpha - \Theta/2) \right\rangle 
  \pm(-1)^n \left \langle  (\widehat{x}_j^{\pm}(\alpha-\Theta/2))^\dag x_j^{[r]}(\alpha + \Theta/2) \right\rangle \right)
 \right|_{\alpha= \beta=0}
 \nonumber \\
 \label{eq:qq}
\eea 

From the previous analysis, (\ref{eq:xx}) involves a product of $\sim r/2$ factors of the form $\mathcal{C} \left( \substack{ \alpha + \Theta/2 \\ \alpha +\Theta/2 } \right)\mathcal{C} \left( \substack{ \alpha-\Theta/2  \\ \alpha-\Theta/2  } \right)$, so for large $r$ it decreases as  
\be 
|| x_j^{ [r]}(\alpha)||_{\rm HS}^2 \sim  \left(  f_j(\alpha + \Theta/2) f_j(\alpha- \Theta/2) \right)^{r \over 2}   
\sim \mathrm{e}^{- r/ \xi_j(\alpha - \Theta/2)}
 \,,
 \nonumber
\ee
 where 
\be
	\xi_j(\alpha)  = -{2 \over   \log  \left(f_j(\alpha) f_j(\alpha + \Theta))\right) }  \,.
	\label{eq:xij}
\ee

Turning to (\ref{eq:qq}), the four terms on the right hand side respectively a product of $\sim r/2$ factors of the form $\mathcal{C} \left( \substack{  \Theta \\ \Theta } \right)\mathcal{C} \left( \substack{ 0 \\0  } \right)$,  $\mathcal{C} \left( \substack{0 \\ 0 } \right)\mathcal{C} \left( \substack{ -\Theta \\  -\Theta } \right)$,
 $\mathcal{C} \left( \substack{  \Theta \\ 0} \right)\mathcal{C} \left( \substack{0\\  -\Theta } \right)$, 
 $\mathcal{C} \left( \substack{ 0  \\  - \Theta} \right)\mathcal{C} \left( \substack{  \Theta\\ 0 } \right)$. From the properties listed above the two first terms dominate in the large $r$ limit, and one has therefore 
\be 
|| q_j^{\pm[r]}||_{\rm HS}^2 \sim  \left(  f_j(\Theta) f_j(0) \right)^{r \over 2} 
\sim \mathrm{e}^{- r/ \xi_j(0)}
 \,.
 \nonumber
\ee

The scaling limit of the light-cone lattice is defined (see equation (\ref{eq:scalinglimit})) by taking $\Lambda \to \infty$ (so $\Theta \to \infty$), while keeping $\alpha$ finite. In this limit one has 
\be
\xi_j(\alpha)  = -{2 \over   \log  \left(f_j(\alpha )\right) }  \,,
\nonumber
\ee
 which can be accessed through the numerical knowledge of  the functions $a(j,\gamma)$ and $b(j,\gamma)$ in (\ref{eq:ajbj}). 
In the figure \ref{fig:correlationlengths} of the main text, we display plots of the correlation lengths $\xi_j$ for $\alpha$ in the physical domain and various values of $j$, as a function of $\gamma$. Fron this figure it is concluded (see main text for details) that the correlation length $\xi_j$ is finite for $\gamma < \frac{\pi}{j}$, and diverges otherwise, which gives an alternative illustration of the fact noticed in \cite{Ilievski-stringcharge}, namely that the quasilocal charge content is intimately related to the string content of the model.

\section{Previously known fractional spin charges in sine-Gordon}
\label{app:othercharges}

\subsection{The Bernard-LeClair quantum group charges}

The earliest example of fractional spin charges in sine-Gordon of which we are aware are the generators of quantum group symmetries discovered by Bernard and Leclair in \cite{Bernard1,Bernard2}. These charges can be easily built by starting with the massless field of a bosonic CFT that is the ultra-violet limit of sine-Gordon theory. In the ultraviolet CFT, one can parametrize spacetime using the complex coordinates $w,\bar{w}$, and the bosonic sine-Gordon field separates into holomorphic and antiholomorphic components:
\bea
\varphi(w,\bar{w})=\phi(w)+\bar{\phi}(\bar{w}).
\nonumber
\eea
It is also convenient to define the field 
\bea
\Theta(w,\bar{w})=\phi(w)-\bar{\phi}(\bar{w}),
\nonumber
\eea
which satisfies $\partial_\mu \varphi=-i\epsilon_{\mu\nu}\partial_\nu \Theta$. Given this relationship it is easy to see that the field $\Theta$ is highly non-local relative to $\varphi$ if we switch back to  coordinates $x,t$, and integrate to find the expression
\bea
\varphi(x,t)=\int_{-\infty}^x i \partial_t \theta(y,t)dy,\nonumber
\eea
which explicitly shows that the relation between the two operators involves integrating over all spatial points from minus infinity to $x$. This spatial integration is called a "tail" in the main text.

A main finding of \cite{Bernard1,Bernard2} is that one can define a set of currents, satisfying the conservation equations
\bea
\partial_{\bar{w}}J_{\pm}=\partial_{w}H_{\pm};\,\,\,\partial_w \bar{J}_{\pm}=\partial_{\bar{w}}\bar{H}_{\pm},
\nonumber
\eea
which can be expressed in terms of the fields as
\bea
&J_{\pm}\sim\exp\left(\pm\frac{2 i}{b}\phi\right)=\exp\left(\pm \frac{i}{b}\varphi\pm\frac{i}{b}\Theta\right),\nonumber\\
&H_{\pm}\sim \exp\left[\pm\left(\frac{2}{b}-b\right)\phi\mp i b\bar{\phi}\right]=\exp\left[\pm i\left(\frac{1}{b}-b\right)\varphi\pm \frac{i}{b}\Theta\right],\nonumber\\
&\bar{J}_{\pm}\sim \exp\left(\mp\frac{2 i}{b}\bar{\phi}\right)=\exp\left(\mp\frac{i}{b}\varphi\pm\frac{i}{b}\Theta\right),\nonumber\\
&\bar{H}_{\pm}\sim\exp\left[\mp i\left(\frac{2}{b}-b\right)\bar{\phi}\pm i b\phi\right]=\exp\left[\pm i\left(\frac{1}{b}-b\right)\varphi\pm\frac{i}{b}\Theta\right].\label{non-localcurrents}
\eea
where in our notation, 
\bea
\frac{1}{b^2}=\frac{\gamma}{\pi-\gamma}+\frac{1}{2}.\nonumber
\eea
The fact that these currents are expressed in terms of the non-local fields $\Theta$, implies that they are non-local, as well as the associated conserved charges. From these expressions of the currents, it is easy to find their Lorentz spin, from the fact that the spin of a general vertex operator
\bea
V_{\alpha,\beta}=\exp\left(i\alpha\phi+i\beta\bar{\phi}\right), \nonumber
\eea
is known to be $s=\left(\alpha^2-\beta^2\right)/2$.
Conserved charges are obtained by integrating the currents (\ref{non-localcurrents}), as
\bea
Q_{\pm}=\frac{1}{2\pi i}\left(\int dw J_{\pm}+\int d\bar{w}H_{\pm}\right),\,\,\,\bar{Q}_{\pm}=\frac{1}{2\pi i}\left(\int d\bar{w}\bar{J}_{\pm}+\int dw\bar{H}_{\pm}\right)\,, \label{non-localcharges}
\eea
and their Lorentz spin is
\be s=\frac{\pm 2\gamma}{\pi-\gamma} \,.
\ee

\subsection{The Bazhanov-Lukyanov-Zamolodchikov commuting charges}

We here review some aspects of the construction of Bazhanov, Lukyanov and Zamolodchikov in \cite{Russians,Russians2}.
Given a certain CFT, with central charge, $c$ (initially restricted to $c<-2$, though an analytic continuation for the expressions of the conserved charges was proposed in \cite{Russians2}), and holomorphic and antiholomorphic components of the stress energy tensor, $T(w)$ and $\bar{T}(\bar{w})$, respectively, the starting point of \cite{Russians} is the Feigin-Fuchs free field representation \cite{Fuchs}
\bea
-g^2 T(w)=:\,(\psi^\prime(w))^2\,:+(1-g^2)\psi^{\prime\prime}(w)+\frac{g^2}{24},\,\,\,\,\,\,\,\,\,c=13-6(g^2+g^{-2})\nonumber
\eea
where $\psi(w)$ is a free field, which can be expanded as
\bea
\psi(w)=iQ+iPw+\sum_{n\neq 0}\frac{a_{-n}}{n}e^{inw},\nonumber
\eea
and $:\,\cdot\,:$ represents the normal ordering in terms of the oscillators, $a_n$. The operators $Q,\,P$ and $\{a_n\}_{n\neq0}$ satisfy the algebra
\bea
[Q,P]=\frac{i}{2}g^2,\,\,\,\,[a_n,a_m]=\frac{n}{2}g^2\delta_{n+m,0}.\nonumber
\eea
A similar expansion can be done for $\bar{T}(\bar{w})$ in terms of $\bar{\psi}(\bar{w})$.

The next step in \cite{Russians} is to construct a representation for the transfer matrices $T_j$, which is written in terms of the elements of the quantum algebra, $U_q(sl(2))$, with $q=\exp(i\pi g^2)$, and the vertex operators,
\bea
V_{\pm}(w)=:\,e^{\pm 2 \psi(w)}\,:,\nonumber
\eea
which with the normalization conventions of \cite{Russians}, have conformal dimension, $\Delta=g^2$. A set of non-local conserved charges is found from expanding the transfer matrices in powers of the spectral parameter. For the transfer matrix corresponding to the spin $1/2$ representation of the quantum group, the non-local conserved charges can be expressed as (for $c<-2$)
\bea
G_{2n}=q^{n}\int_{w_1\geq\cdots\geq w_{2n}}^{2\pi} && dw_1\cdots dw_{2n} e^{2\pi i P} V_{-}(w_1)V_{+}(w_2)\cdots V_{-}(w_{2n-1})V_{+}(w_{2n})\nonumber\\
&&+e^{-2\pi i P}V_{+}(w_1)V_{-}(w_2)\cdots V_+(w_{2n-1})V_{-}(w_{2n}),\label{chargesBLZ}
\eea
for integers, $n$.  For more generic values of $c$, the expression (\ref{chargesBLZ}) can be generalized by exchanging the ordered integrals with contour integrals that do not diverge (as is shown in Eq. 2.19 of \cite{Russians2}).

The charges (\ref{chargesBLZ}) are manifestly non-local, as attested from the fact that the given expression involves many spatial integrals connecting distant points in space. One immediate peculiarity we can notice is that these charges have in general fractional spin. This is seen, similarly as for the charges (\ref{non-localcharges}), from the fact that the integrand in (\ref{chargesBLZ}) is expressed in terms of vertex operators whose spin is known, and $g$-dependent. From this argument, one can read that the Lorentz spin of the charge (\ref{chargesBLZ}) is $s=2n(g^2-1)$.

The connection between these conserved charges of CFT and  those of sine-Gordon is explained in more detail in Ref. \cite{negro}. This consists of two steps:  first, define the transfer matrices for massive integrable deformations of the CFT's. In the massive case, the transfer matrices involving the holomorphic and antiholomorphic components of the Feigin-Fuchs field are not independent, but one set of transfer matrices is obtained by joining both components.
The second step is to use the fact that the $\Phi_{1,3}$ deformations of unitary minimal models can be obtained from a restricted sine-Gordon theory, which allows one to relate the sine-Gordon field to the Feigin-Fuchs field. 

As was shown in Ref. \cite{Bernard1}, the connection between sine-Gordon and perturbed minimal models comes from the fact that the solitons and antisolitons of SG transform as a two-dimensional representation of the quantum group $U_{\tilde{q}}(sl(2))$ with $\tilde{q}=\exp(-i2\pi/b^2)$. At infinite volume, the Hilbert space of sine-Gordon contains subspaces which are annihilated by different elements of $U_{\tilde{q}}(sl(2))$, these subspaces are then associated with the Hilbert space of the perturbed minimal models. The identification of the restricted sine-Gordon with the perturbed minimal models leads to the association $\tilde{q}=\exp(i\pi g^2)=\exp(i 2\pi/b^2)$, or $g^2=2/b^2$. Using this identification and the definition of $b$, we find that for sine-Gordon, the non-local conserved charges, $G_{2n}$ have Lorentz spin
\be 
s=\frac{\pm 2\gamma n}{\pi-\gamma} \,.
\label{spinBLZ}
\ee

\end{appendix}

\end{document}